\documentclass[11pt,preprint]{aastex}

\usepackage{graphicx,amsmath}
\usepackage{color}
\usepackage[plainpages=false, colorlinks=true, anchorcolor=blue,
linkcolor=blue, citecolor=blue, bookmarks=false]{hyperref}

\newcommand{\jump}[1]{\Delta_s \left( #1 \right)}
\newcommand{\Msun}{{\rm\,M_\odot}}
\newcommand\kms{{\rm\, km\, s^{-1}}}
\newcommand\pc{{\rm\, pc}}
\newcommand\kpc{{\rm\, kpc}}
\newcommand\Gyr{{\rm\,Gyr}}
\newcommand\cs{c_s}
\newcommand\alfv{v_{\rm A0}}
\newcommand\alfc{v_{{\rm A},c}}
\newcommand\BBeta{\beta}
\newcommand\cA{\alpha_0}

\newcommand\F{\mathcal{F}}
\newcommand\vf{\mathbf{v}}
\newcommand\Bf{\mathbf{B}}
\newcommand\Bfi{\mathbf{B}_0}

\newcommand\omgD{\omega_D}

\newcommand{\omgR}{{\rm Re}(\omega)}
\newcommand{\omgI}{{\rm Im}(\omega)}
\newcommand\uT{u_{T0}}
\newcommand\vT{v_{T0}}

\newcommand\spost{{s+}}
\newcommand\spre{{s-}}
\newcommand\DX{\eta}
\newcommand\mach{\mathcal{M}}
\newcommand\xSO{x_{\rm mp}}
\newcommand\xSH{x_{\rm sh}}
\newcommand\kx{k_x}

\newcommand\ky{k_y}

\newcommand\freq{{\rm\, km\, s^{-1}\, kpc^{-1}}}

\newcommand\simgt{\lower.5ex\hbox{$\; \buildrel > \over \sim \;$}}
\newcommand\simlt{\lower.5ex\hbox{$\; \buildrel < \over \sim \;$}}

\newcommand\mQ{\mathcal{E}}
\newcommand\be{\begin{equation}}
\newcommand\ee{\end{equation}}
\newcommand\bea{\begin{eqnarray}}
\newcommand\eea{\end{eqnarray}}

\shorttitle{Wiggle Instability of Magnetized Galactic Spiral Shocks} %
\shortauthors{Kim, Kim, \& Elmegreen}

\begin{document}

\title{Wiggle Instability of Galactic Spiral Shocks: Effects of Magnetic
Fields}\author{Yonghwi Kim$^{1}$, Woong-Tae Kim$^1$, and Bruce G.\
Elmegreen$^2$}

\affil{$^1$Center for the Exploration of the Origin of the Universe
(CEOU), Astronomy Program, Department of Physics \& Astronomy, Seoul
National University, Seoul 151-742, Republic of Korea \\
$^2$IBM T. J. Watson Research Center, 1101 Kitchawan Road, Yorktown
Heights, New York 10598, USA}

\email{kimyh@astro.snu.ac.kr, wkim@astro.snu.ac.kr, bge@us.ibm.com}

\begin{abstract}
It has been suggested that the wiggle instability (WI) of spiral shocks
in a galactic disk is responsible for the formation of gaseous feathers
observed in grand-design spiral galaxies. We perform both a linear
stability analysis and numerical simulations to investigate the effect
of magnetic fields on the WI. The disk is assumed to be
infinitesimally-thin, isothermal, and non-self-gravitating.  We control
the strengths of magnetic fields and spiral-arm forcing using the
dimensionless parameters $\beta$ and $\F$, respectively. By solving the
perturbation equations as a boundary-eigenvalue problem, we obtain
dispersion relations of the WI for various values of
$\beta=1$--$\infty$ and $\F=5\%$ and 10\%. We find that the WI arising
from the accumulation of potential vorticity at disturbed shocks is
suppressed, albeit not completely, by magnetic fields. The stabilizing
effect of magnetic fields is not from the perturbed fields but from the
unperturbed fields that reduce the density compression factor in the
background shocks.  When $\F = 5\%$ and $\beta\lesssim 10$ or $\F=10\%$
and $\beta\sim5$--10, the most unstable mode has a wavelength of
$\sim0.1$--0.2 times the arm-to-arm separation, which appears
consistent with a mean spacing of observed feathers.
\end{abstract}


\section{Introduction}

Stellar spiral arms in disk galaxies greatly affect galaxy
evolution in various ways (e.g., \citealt{but96,kor04,but13,sel14}).
They not only trigger and/or organize star formation by
compressing gas into spiral shocks (e.g.,
\citealt{rob69,rob70,shu72,shu73}) but also drive secular galaxy
evolution (e.g., \citealt{lin64,lin66,too64,elm95,ber96,foy10}).
Observations commonly indicate that grand-design spiral arms abound
with secondary structures called stellar ``spurs'' or gaseous/dust
``feathers'' that stretch out almost perpendicularly from the main arms
into the interarm regions in a trailing configuration (e.g.,
\citealt{baa63,lyn70,wea70,elm83,sco01a,sco01b,
ken04,wil04,lav06,cor08,sil12,sch13}). While spurs and feathers have
similar pitch angles and thus are thought to share the common origin
(\citealt{elm80}; see also \citealt{pue14}), what actually forms them
has been a matter of considerable debate.

Existing theories for the formation of secondary structures differ in
the relative role played by gaseous self-gravity to other hydrodynamic
processes. \citet{bal85} and \citet{bal88} studied a linear
stability of spiral shocks to axisymmetric and non-axisymmetric modes,
respectively, and showed that these arm substructures may form as a
consequence of swing amplification of local density perturbations.
\citet{elm94} showed that local perturbations can grow faster in the
presence of magnetic fields that can remove the constraint of angular
momentum conservation. These linear-theory predictions were confirmed
by numerical simulations \citep{kim02a,kim06,she06} that showed that
both magnetic fields and self-gravity are essential to form secondary
structures resembling feathers in the nonlinear stage. Recently,
\citet{lee12} treated a stability of self-gravitating, magnetized
spiral shock as a global, rather than local, problem in the direction
perpendicular to the arm. They neglected galactic shear for
perturbations while keeping non-inertial terms, and found
semi-analytically that such shocks are prone to feather formation.
Although these feather-forming instabilities are termed differently as
the azimuthal instability, magneto-Jeans instability, and feathering
instability by \citet{elm94}, \citet{kim02a}, and \citet{lee12},
respectively, they refer to the same gravitational instability of a
rotating medium in which embedded magnetic fields play a destabilizing
role. More recently, \citet{lee14} extended the work of \citet{lee12}
to explore the parameter space of the feathering instability by varying
the strength of magnetic fields and self-gravity.

In contrast, other studies argued that gaseous self-gravity may not be
indispensable for the formation of feathers (e.g.,
\citealt{joh86,wad04,dob06,dob07}).  These authors showed that
secondary structures develop even in the absence of self-gravity, with
their spacings smaller than those resulting from the self-gravitating
instabilities mentioned above. In particular, \citet{wad04} observed
that small clumps form in strongly shocked layers that may grow into
feathers, and termed the clump-forming mechanism the wiggle instability
(WI). The WI appears prevailing in recent numerical simulations of
spiral galaxies \citep{kim14b} and even barred galaxies with strong
dust-lane shocks \citep{kim12a,kim12b,kimsto12,seo13,seo14}. Based on
the Richardson number criterion (e.g, \citealt{cha61}), they further
proposed that the WI is originated from the Kelvin-Helmholtz
instability (KHI) of a shear layer behind the shock. However, the
linear analysis of \citet{dwa96} showed that postshock flows in the
presence of rapidly varying shear are stable to the KHI. Interestingly,
\citet{han12} argued that the WI may result spuriously from the
difficulty in resolving a shock inclined to numerical grids.

To address the nature of the WI, \citet[hereafter Paper~I]{kim14a}
adopted the method of \citet{lee12} and performed a linear stability
analysis of non-self-gravitating, unmagnetized spiral shocks. Paper~I
found that the WI is physical rather than numerical, arising from the
generation of potential vorticity (PV) at a distorted shock front,
known as Crocco's theorem.  Since gas in galaxy rotation passes through
spiral shocks twice in every rotation for two-armed spirals, the PV
contained in entropy-vortex waves can increase successively, in a
Lagrangian sense, in the course of galaxy rotation. This sets up an
Eulerian overstable normal mode along the azimuthal direction that
grows exponentially. This is quite distinct from the KHI that relies on
shear in the background flows. By relying on the growth of
incompressible entropy-vortex modes, the WI is also different from the
magneto-Jeans or feathering instability mentioned above that utilizes
compressive acoustic modes. Paper~I also confirmed that the growth
rate and wavelength of the most unstable mode found by the linear
stability analysis are consistent with the results of direct numerical
simulations.

While the results of Paper~I are informative to understand the nature
of the WI, they are based on the models that do not consider magnetic
fields pervasive in the interstellar medium (ISM). It has been well
recognized that disk galaxies have a large-scale, ordered component as
well as a small-scale, random component of magnetic fields (e.g.,
\citealt{wie93,hei95,bec96,bec01}). In most disk galaxies, magnetic
field directions based on polarized synchrotron radiation follow
optical spiral structures fairly well, with their strength and pitch
angle ranging typically $\sim 4-20\,\mu$G and $\sim 8^\circ-37^\circ$,
respectively (see e.g., \citealt{nei92, bec96,van15}). Although there
is an indication that dust lanes upstream of optical spiral arms have
strongest magnetic fields, some regular fields also extend well into
interarm regions. It appears that small scale activities such as star
formation and supernova explosions inside spiral arms disrupt regular
magnetic fields and turn them into turbulent ones. When compressed by
spiral shocks, these turbulent fields become as strong as, sometimes
even stronger than, the regular component inside the arms (e.g.,
\citealt{fle11,hou13,shn14}). For gas surface densities of
$\Sigma\sim10-100\Msun\pc^{-2}$ inside the arms, these fields are close
to equipartition strength with the thermal and turbulence energies,
indicating that they are important in the gas dynamics associated with
the arms (e.g., \citealt{bec96,chy00,bec07}).

In this paper, we extend Paper~I by including the effects of magnetic
fields on the stability of spiral shocks.  As in Paper~I, we consider a
local shearing sheet of an infinitesimally-thin gaseous disk that is
assumed to be isothermal and non-self-gravitating. To simplify the
situation further, we consider only the regular component of magnetic
fields that is initially parallel to the imposed stellar spiral arms.
One-dimensional (1D) solutions of equilibrium magnetized spiral
shocks were already obtained by several authors (e.g.,
\citealt{rob70,kim02a,lee14}). By imposing two-dimensional (2D)
perturbations onto such shocks, we will show that the presence of
magnetic fields substantially suppresses the development of the WI
mainly by reducing the density compression factor of a background
spiral shock. The tension and pressure forces from the perturbed
magnetic fields will turn out to have a minor effect on the WI. The
stabilizing effect of magnetic fields is qualitatively consistent with
the results of the previous magnetohydrodynamic (MHD) simulations
\citep{kim06,she06,dob08}. We also run local MHD simulations to confirm
the results of our linear stability analysis, and compare the pitch
angles of the nonlinear structures formed in the simulations with the
values reported by \citet{pue14} for observed feathers in the
grand-design spiral galaxies M51 and M81.

This paper is organized as follows. In Section \ref{sec:steady}, we
introduce the basic MHD equations and our choices of the parameters, as
well as the equilibrium shock solutions we use as a background state of
the WI. In Section \ref{sec:method}, we derive the linearized
perturbation equations for our normal-mode stability analysis and
present the shock jump conditions. We also give the computational
method to obtain the dispersion relations. In Section \ref{sec:disp},
we present the resulting dispersion relations for 1D and 2D modes as
well as the results of the MHD simulations, and discuss the effect of
magnetic fields on the WI. In Section \ref{sec:summary}, we summarize
our findings and discuss their astronomical implications.

\section{Steady Spiral Shocks}\label{sec:steady}

\subsection{Basic Equations}\label{sec:beq}

We consider magnetized spiral shocks in a galactic disk and study their
stability to small-amplitude perturbations. The disk is assumed to be
infinitesimally thin, isothermal with the speed of sound $\cs$, and
non-self-gravitating: the effect of self-gravity will be studied in a
separate work.

Following \citet{rob69}, we employ a local Cartesian coordinate system
$(x, y)$ with the $x$- and $y$-axes representing the directions
perpendicular and parallel to a local segment of a spiral arm,
respectively (see also \citealt{shu73,bal88}). We place the local frame
at the galactocentric distance $R$ and let it corotate with the arm at
the pattern speed $\Omega_p$. Let $i$ denote the pitch angle of the arm
segment. By taking the local approximation ($|x|, |y| \ll R$) and
assuming that the arm is tightly wound ($\sin i\ll 1$), the background
gas velocity in the local frame arising from galaxy rotation is given
by
 \be\label{e:bgvel}
  \mathbf{v}_c \equiv u_c \hat{\mathbf{x}} + v_c \hat{\mathbf{y}}
  = R(\Omega-\Omega_p)\sin i\,
  \hat{\mathbf{x}}+[R(\Omega-\Omega_p)-q_c\Omega x]\,\hat{\mathbf{y}},
 \ee
where $\Omega$ is the rotational angular velocity at $R$ and
$q_c\equiv-d\ln\Omega/d\ln R$ is the local shear rate in the absence of
the arm: the term involving $q_c$ handles differential rotation in our
local models. The corresponding epicycle frequency is $\kappa =
(R^{-3}d[\Omega^2 R^4]/dR)^{1/2}=(4-2q_c)^{1/2}\Omega$. Note that
$\mathbf{v}_c$ in Equation \eqref{e:bgvel} depends only on $x$ and is
independent of $y$ (see also \citealt{kim02a}), which allows to explore
the behaviors of periodic waves in the $y$-direction.

Since the length scales involved in gas dynamics over galactic
disks are enormously large compared to electrical diffusion scales
(\citealt{rob70}; see also \citealt{shu92}), the magnetic Reynolds
number is much larger than unity. In this case, one can make the
``frozen-in-field'' approximation in which the ISM is assumed to be
tightly coupled to the magnetic fields. Expanding the compressible,
ideal MHD equations in the local frame, one obtains
 \be\label{e:con}
  \frac{\partial\Sigma}{\partial t}+\nabla\cdot(\Sigma\vf_T)=0,
 \ee
 \be\label{e:mom}
  \frac{\partial\vf}{\partial t} + \vf_T\cdot\nabla\vf
  =-\cs^2\nabla\ln\Sigma+q_c\Omega v_x\hat{\mathbf{y}}
  -2\mathbf{\Omega}\times\vf-\nabla\Phi_s
  +\frac{1}{4\pi\Sigma}(\nabla\times\Bf)\times\Bf,
 \ee
 \be\label{e:ind}
  \frac{\partial\Bf}{\partial t}=\nabla\times(\vf_T\times\Bf),
 \ee
together with
 \be\label{e:dfree}
   \nabla\cdot\Bf =0,
 \ee
where $\Sigma=\int \rho dz$ is the gas surface density, $\vf$ is the
gas velocity induced by the arms, and $\vf_T=\vf+\vf_c$ is the total
gas velocity in the local frame. In Equations \eqref{e:mom} and
\eqref{e:ind}, $\mathbf{B}=H^{1/2}\mathbf{B}_{\rm 3D}$, where
$\mathbf{B}_{\rm 3D}$ is the midplane value of the three-dimensional
magnetic fields and $H=\Sigma/\rho$ is the disk thickness that is
assumed to be constant over $R$ (e.g., \citealt{kim01,kim02a}).

In Equation \eqref{e:mom}, $\Phi_s$ denotes the gravitational potential
of the stellar spiral arms, for which we take a simple sinusoidal
shape:
 \be\label{e:extP}
  \Phi_s = \Phi_0 \cos \left(\frac{2\pi x}{L}\right),
 \ee
constant along the $y$-direction. Here, $\Phi_0 (<0)$ is the amplitude
and $L =2\pi R\sin i /m$ is the arm-to-arm distance for $m$-armed
spirals. We consider a domain with $-1/2 \leq x/L \leq 1/2$, so
that the spiral potential achieves its minimum at the domain center
($x=0$). The arm strength can be characterized by the dimensionless
parameter
 \be\label{e:F}
  \F\equiv \frac{m}{\sin i} \left(\frac{|\Phi_0|}{R^2\Omega^2}\right),
 \ee
which corresponds to the maximum gravitational force due to the spiral
arms relative to the centrifugal force of the background galaxy
rotation (e.g., \citealt{rob69}).

In the absence of spiral arms, the gaseous disk has a uniform surface
density $\Sigma_c$ and uniform thickness-normalized magnetic
fields $B_c\hat{\mathbf{y}}$ parallel to the arms. We quantify the
strength of the magnetic fields using the  plasma parameter
 \be\label{e:betac}
 \BBeta
 \equiv \frac{\cs^2}{\alfc^2} = \frac{4\pi\cs^2\Sigma_c}{B_c^2}
 =\frac{4\pi P_c}{B_{{\rm 3D},c}^2}\,,
 \ee
where $\alfc^2\equiv B_c^2/4\pi\Sigma_c$ is the Alfv\'en speed, and
$P_c=\cs^2\rho_c$ is the mean thermal pressure of the ISM at the
galactic plane. Adopting the fiducial values $P_c/k_B\sim2000-3000$ K
cm$^{-3}$ \citep{hei01} and $B_{{\rm 3D},c}=1.4\mu\rm G$ \citep{ran94}
in the solar neighborhood, $\BBeta\simeq 4$, but we consider a range of
$\BBeta$ to study various situations with differing field strength.

In Appendix \ref{a:vor}, we combine Equations
\eqref{e:con}--\eqref{e:dfree} to obtain
 \be\label{e:pvcon}
  \left(\frac{\partial }{\partial t}+\vf_T\cdot\nabla \right)
  \boldsymbol{\xi} = \frac{\mathbf{B}\cdot\nabla}{4\pi\Sigma}
  \left(\frac{\nabla\times\mathbf{B}}{\Sigma}\right),
 \ee
where
 \be\label{e:pv}
  \boldsymbol{\xi} \equiv \frac{\nabla\times\vf_T+2
  \boldsymbol{\Omega}}{\Sigma}
 \ee
is PV.  This states that PV is, in general, not conserved in the
presence of magnetic fields. This is unlike in 2D unmagnetized flows
where $\boldsymbol{\xi}$ remains constant along a given streamline.

Equations \eqref{e:con}--\eqref{e:betac} can be completely specified by
seven parameters: $q_c$, $m$, $\sin i$, $\Omega_p/\Omega$,
$\cs/(R\Omega)$, $\F$, and $\BBeta$. We fix to $q_c = 1$, $m = 2$,
$\sin i = 0.1$, $\Omega_p/\Omega=0.5$, and $\cs/(R\Omega) =0.027$, and
vary $\F= 5\%$--$10\%$ and $\BBeta=1$--$\infty$ for our presentation
below.

\subsection{Equilibrium Shock Profiles}\label{sec:equil}

We now seek for 1D steady solutions of Equations
\eqref{e:con}--\eqref{e:dfree} subject to a spiral-arm forcing with
strength $\F$, which will be used as an unperturbed equilibrium state
in Section \ref{sec:method}. Let us denote the steady solutions using
the subscript ``0'' as $\Sigma_0(x)$,
$\vf_0=u_0(x)\hat{\mathbf{x}}+v_0(x)\hat{\mathbf{y}}$, and
$\Bfi=B_0(x)\hat{\mathbf{y}}$.  Then, the steady-state conditions yield
 \be\label{e:con0}
  \Sigma_0 \uT =  \Sigma_{c} u_c = \text{constant},
 \ee
 \be\label{e:momx0}
  \uT \frac{du_0}{dx} =
  -\frac{c_s^2}{\Sigma_0}\frac{d\Sigma_0}{dx} + 2 \Omega v_0 -
  \frac{d\Phi_s}{dx}-\frac{B_0}{4\pi\Sigma_0}\frac{dB_0}{dx},
 \ee
 \be\label{e:momy0}
  \uT\frac{dv_0}{dx} = -\frac{\kappa^2}{2\Omega} u_0,
 \ee
and
 \be\label{e:ind0}
  B_0 \uT =  B_{c} u_c = \text{constant}.
 \ee
Equations \eqref{e:con0} and \eqref{e:ind0} imply that $B_0 \propto
\Sigma_0$, a usual relation resulting from the conservation of mass and
magnetic flux in one dimension. Eliminating $\Sigma_0$ and $B_0$ in
favor of $\uT$, Equation \eqref{e:momx0} for $m=2$ reduces to
 \be\label{e:momx01}
  \left(\uT-\frac{\cs^2}{\uT}-\frac{\alfv^2}{\uT}\right)\frac{d\uT}{dx}
  = 2\Omega v_0 + R\Omega^2 \F \sin\left(\frac{2\pi x}{L}\right),
 \ee
where
 \be
  \alfv \equiv \frac{B_0}{\sqrt{4\pi\Sigma_0}} =
  \alfc \left(\frac{u_c}{\uT}\right)^{1/2},
 \ee
is the Alfv\'en speed in an equilibrium configuration. Note that
$\alfv \propto B_0^{1/2}\propto\uT^{-1/2}$.

It can be shown that the PV in the equilibrium flows is
 \be\label{e:pv0}
  \xi_0 = \frac{|\nabla\times \mathbf{v}_{T0}+2\mathbf{\Omega}|}{\Sigma_0}
  =\frac{\kappa^2}{2\Omega\Sigma_c},
 \ee
which is constant everywhere even in magnetized spiral shocks. This
results from the fact that an equilibrium shock satisfies
$\nabla\times\Bfi=0$ (Eq.~\eqref{e:pvcon}). Note that Equation
\eqref{e:pv0} requires that the local shear rate in a steady-state
spiral shock should vary as
  \be\label{e:shear}
    q_0 \equiv -\frac{1}{\Omega}\frac{d\vT}{dx} = 2 -
    (2-q_c)\frac{\Sigma_0}{\Sigma_c},
  \ee
indicating that shear is reversed wherever $\Sigma_0/\Sigma_c \geq
2/(2-q_c)=2$ for $q_c=1$ \citep{bal85,kim02a}.

The equilibrium velocity profiles can be obtained by solving Equations
\eqref{e:momy0} and \eqref{e:momx01} simultaneously.  We follow
the method given by \citet{shu73} (see also \citealt{lee12}; Paper~I).
The detailed procedure is provided in Appendix \ref{a:1d}. Figure
\ref{f:shock} plots exemplary profiles of equilibrium spiral shocks for
$\F=5\%$ and 10\% and $\BBeta=\infty, 10, 3$, and 1. A filled circle
marks the magnetosonic point, $\xSO$, in each shock profile. Table
\ref{t:1d} lists the associated values of $\xSO$, the shock position
$\xSH$, and the preshock and postshock surface densities
$\Sigma_0^\spre$ and $\Sigma_0^\spost$, the density compression factor
defined by
 \be\label{e:djump}
  \mu \equiv \frac{\Sigma_0^\spost}{\Sigma_0^\spre} =
  \frac{\uT^\spre}{\uT^\spost},
 \ee
as well as the preshock Mach number $\mathcal{M}\equiv \uT^\spre/\cs$.
It is apparent that magnetic fields make the shock weaker by providing
magnetic pressure, reducing $\mu$ considerably.  Since $\Sigma_0^\spre$
and $\mathcal{M}$ are insensitive to $\BBeta$ for fixed $\F$, the
reduction of the compression factor due to magnetic fields occurs
primarily by making the shock front move toward the upstream direction.
Note that shear is reversed in the regions behind the shock front where
$\Sigma_0/\Sigma_c\geq 2$.  The degree of shear reversal defined as
$q_0$ in Equation \eqref{e:shear} is larger as the shock becomes
stronger, which makes the structures that develop as a consequence of
the WI more perpendicular to the arms, as will be shown in Section
\ref{sec:num}.

\section{Linear Stability Analysis}\label{sec:method}

\subsection{Perturbation Equation}\label{sec:perturb}

We apply small-amplitude Eulerian perturbations, denoted by $\Sigma_1$,
$u_1$, $v_1$, and $\Bf_1$, to an equilibrium shock found in the
preceding section. We then linearize Equations
\eqref{e:con}--\eqref{e:dfree} to obtain
 \be\label{e:con1}
  \frac{D_0}{Dt}\left(\frac{\Sigma_1}{\Sigma_0}\right)
  + \frac{\partial u_1}{\partial x}
  + \frac{d\ln\Sigma_0}{dx} u_1
  + \frac{\partial v_1}{\partial y}=0,
 \ee
 \be\label{e:mx1}
  \begin{split}
  \frac{D_0 u_1}{D t} + \frac{du_0}{dx} u_1
  + \cs^2 \frac{\partial}{\partial x}\left(\frac{\Sigma_1}{\Sigma_0}\right)
  - 2\Omega v_1  - \frac{B_0}{4\pi\Sigma_0}\left(\frac{\partial^2}{\partial x^2}
  + \frac{\partial^2}{\partial y^2}
  \right) m_1 \\
  - \frac{1}{4\pi\Sigma_0}\frac{dB_0}{dx}\left(\frac{\partial m_1}{\partial x}
  +B_0\frac{\Sigma_1}{\Sigma_0}\right)=0,
  \end{split}
 \ee
 \be\label{e:my1}
  \frac{D_0 v_1}{D t}
  +\cs^2\frac{\partial}{\partial y}\left(\frac{\Sigma_1}{\Sigma_0}\right)
  + \left(\frac{\kappa^2}{2\Omega}\right)\frac{u_c}{\uT}u_1
  -\frac{1}{4\pi\Sigma_0}\frac{dB_0}{dx}\frac{\partial m_1}{\partial y}=0,
 \ee
and
 \be\label{e:ind1}
  \frac{D_0 m_1}{D t} = B_0 u_1,
 \ee
where
  \be\label{e:Lag}
     \frac{D_0}{Dt} =  \frac{\partial }{\partial t} + \uT \frac{\partial }{\partial x}
  + \vT\frac{\partial }{\partial y},
  \ee
and $m_1(x,y,t)$ is the perturbed vector potential defined through
$\Bf_1\equiv\nabla\times(m_1 \hat{\mathbf{z}})$.

Since the coefficients in Equations \eqref{e:con1}--\eqref{e:ind1}
depend only on $x$, we may consider perturbations of the form
\be\label{e:ptb} \left(
  \begin{array}{c}
    \Sigma_1/\Sigma_0 \\ u_1 \\ v_1 \\ m_1/B_c
  \end{array}\right) = \left(
  \begin{array}{c}
    S_1(x)\\ U_1(x) \\ V_1(x) \\ M_1(x)
  \end{array}\right)
  \exp(-i\omega t + ik_yy),
\ee where $\omega$ and $k_y$ are the frequency and $y$-wavenumber of
the perturbations, respectively. Equations
\eqref{e:con1}--\eqref{e:ind1} then reduce to
 \be\label{e:ds1}
  \begin{split}
  (\uT^2-\cs^2-\alfv^2)\frac{dS_1}{dx}
  \,\,=&\,\,\left[i\omgD\left(\uT-\frac{\alfv^2}{\uT}\right)
  +\frac{\alfv^2}{\uT}\frac{du_0}{dx}\right]S_1 \\
  &\,\,+ \left(2\frac{du_0}{dx} - i\omgD \right)
  \left(1 + \frac{\alfv^2}{\uT^2}\right) U_1 
  -\left[ik_y\left(\uT-\frac{\alfv^2}{\uT}\right)+2\Omega\right]V_1\\
  &\,\,+\frac{\alfv^2}{u_c\uT}\left(\omgD^2+2i\omgD
  \frac{du_0}{dx} + ik_y \uT\frac{d\vT}{dx} +k_y^2\uT^2\right)M_1,
  \end{split}
 \ee
 \be\label{e:du1}
  \begin{split}
  (\uT^2-\cs^2-\alfv^2)\frac{dU_1}{dx}
  \,\,=&\,\,-\left(i\cs^2\omgD+\alfv^2\frac{du_0}{dx}\right) S_1
  +\left( ik_y\cs^2+2\uT\Omega\right ) V_1 \\
  &\,\, + \left[ i \omgD \left( \uT + \frac{\alfv^2}{\uT} \right) -
   \left(\uT +  \frac{\cs^2+3\alfv^2}{\uT}  \right) \frac{du_0}{dx} \right]U_1 \\
  &\,\,-\frac{\alfv^2}{u_c}\left(\omgD^2
  +2i\omgD\frac{du_0}{dx}+ ik_y \uT\frac{d\vT}{dx} + k_y^2\uT^2\right)M_1,
  \end{split}
 \ee
 \be\label{e:dv1}
  \uT\frac{dV_1}{dx}=-ik_y\cs^2 S_1-\frac{\kappa^2}{2\Omega}\frac{u_c}{\uT}
  U_1+i\omgD V_1- i k_y \frac{\alfv^2}{u_c}\frac{du_0}{dx} M_1,
 \ee
and
 \be\label{e:db1}
  \uT\frac{dM_1}{dx}=i\omgD M_1+\frac{u_c}{\uT}U_1,
\ee
where
\be\label{e:dOm}
  \omgD (x) = \omega - k_y \vT
\ee is the Doppler-shifted frequency. We take a convention that $\ky$
is a pure real number and $\omega$ is a complex number.

\subsection{Shock Jump Conditions}\label{sec:jump}

The applied perturbations also disturb the shock front into a
sinusoidal shape. Let the shape of the perturbed shock front be
described by
 \be\label{e:ps}
  \zeta_1 (x,y,t) =  Z_1 \exp (-i\omega t + ik_yy),
 \ee
with the constant amplitude $Z_1$. The unit vectors normal and
tangential to the instantaneous shock front are given by $\hat{\bf
n}=(1,-ik_y\zeta_1)$ and $\hat{\bf t}=(ik_y\zeta_1, 1)$, respectively,
while the velocity of the shock front is ${\bf v}_{\rm
sh}=(-i\omega\zeta_1,0)$ to the first order in $\zeta_1$
(\citealt{dwa96,lee12}; Paper I). It is then straightforward to show
that the perturbations at the perturbed shock positions can be written
as
 \begin{mathletters}\label{e:ttt}
 \bea
  \Sigma (\xSH + \zeta_1) &\approx & \Sigma_0 + \Sigma_1
  + \zeta_1\frac{d\Sigma_0}{dx},\label{e:ts} \\
  v_\perp (\xSH + \zeta_1) &\approx& \uT + u_1 +\zeta_1 \frac{d\uT}{dx}
  +i\omgD \zeta_1,\label{e:tpp} \\
  v_\parallel (\xSH + \zeta_1) &\approx& \vT + v_1
  +\zeta_1 \frac{d\vT}{dx} + ik_y\zeta_1\uT,\label{e:tpr} \\
  B_\perp (\xSH + \zeta_1) &\approx& ik_y  m_1
  -ik_y\zeta_1 B_0,\label{e:tbp} \\
  B_\parallel(\xSH + \zeta_1) &\approx& B_0 - \frac{\partial m_1}{\partial x}
  + \zeta_1 \frac{dB_0}{dx},\label{e:tbr}
 \eea
 \end{mathletters}
where the ``$\perp$'' and ``$\parallel$'' signs denote the components
perpendicular and parallel to the instantaneous shock front in the
stationary shock frame, respectively. All the quantities in the
right-hand side of Equation \eqref{e:ttt} are evaluated at $x=\xSH$.

The jump conditions at the perturbed shock location are given by
\begin{mathletters}\label{e:rk}
\bea
  \jump{ v_\perp \Sigma } &=& 0, \label{e:rk1} \\
  \jump{(\cs^2 + v_\perp^2)\Sigma-\frac{B_\perp^2-B_\parallel^2}{8\pi}}
  &=& 0,\label{e:rk2} \\
  \jump{\Sigma v_\perp v_\parallel -\frac{B_\perp B_\parallel}{4\pi}}
  &=& 0,\label{e:rk3} \\
  \jump{B_\perp v_\parallel - B_\parallel v_\perp} &=& 0, \label{e:rk4} \\
  \jump{B_\perp} &=& 0, \label{e:rk5}
\eea
\end{mathletters}
(e.g., \citealt{shu92}). Substituting Equation \eqref{e:ttt} into
Equation \eqref{e:rk}, one can see that the zeroth-order terms results
in Equation \eqref{e:sj}. The first-order terms are grouped to yield
 \be\label{e:j1}
 \Sigma_0\uT \jump{S_1} + \jump{\Sigma_0 U_1} + iZ_1\omgD^s
  \jump{\Sigma_0}=0,
 \ee
 \be\label{e:j2}
  \begin{split}
  \left(\frac{\uT^2+\cs^2}{2\uT}\right)\jump{S_1}
  +\left(1-\frac{\alfv^2}{2\uT^2}\right)\jump{U_1}
  -i\omgD^s\frac{\cs^2}{2\uT^2\BBeta}\jump{M_1} \\
  +Z_1\Delta_s\left[\left(\frac{\uT^2-\cs^2-\alfv^2}{2\uT^2}\right)
  \frac{d\uT}{dx}\right]=0,
  \end{split}
 \ee
 \be\label{e:j3}
  \jump{V_1}-Z_1\left(\frac{\kappa^2}{2\Omega}\frac{u_c}{\cs^2}
  -i\ky\right)\jump{\uT}+ik_y Z_1\jump{\frac{\alfv^2}{\uT}}
  -ik_y\frac{\cs^2}{\BBeta}\Delta_s\left(\frac{M_1}{\uT}\right) =0,
 \ee
and
 \be\label{e:j4}
  \jump{M_1} -Z_1\jump{B_0}/B_c=0,
 \ee
where $\omgD^s=\omgD(\xSH)$. Note that Equations \eqref{e:rk4} and
\eqref{e:rk5} give the same result, Equation \eqref{e:j4}.

\subsection{Expansion near the Magnetosonic Point}\label{sec:magsonic}

Since the left-hand sides of Equations \eqref{e:ds1} and \eqref{e:du1}
become identically zero at the magnetosonic point, the right-hand sides
should also vanish at $x=\xSO$ in order for regular solutions to exist.
Let us expand the perturbation variables near $x=\xSO$ as
\begin{mathletters}\label{e:exp}
\bea
  S_1 &=& a_0+a_1\DX+\mathcal{O}(\DX^2), \label{e:exp1} \\
  U_1 &=& b_0+b_1\DX+\mathcal{O}(\DX^2), \label{e:exp2} \\
  V_1 &=& c_0+c_1\DX+\mathcal{O}(\DX^2), \label{e:exp3} \\
  M_1 &=& d_0+d_1\DX+\mathcal{O}(\DX^2), \label{e:exp4}
\eea
\end{mathletters}
with coefficients $a_{0,1}$, $b_{0,1}$, $c_{0,1}$, and $d_{0,1}$. 
Substituting Equations \eqref{e:so0} and \eqref{e:exp} into Equations
\eqref{e:ds1}--\eqref{e:db1} and taking zeroth- and first-order terms
in $\DX$, one can obtain a system of five linear equations for the
coefficients.  This indicates that the solutions near $x=\xSO$ can be
completely specified by three constants $a_0$, $b_0$ and $d_0$. Since
the resulting equations are not illuminating, we do not present them
here.

As in Paper I, we solve Equations \eqref{e:ds1}--\eqref{e:db1} as a
boundary value problem with eigenvector $(S_1, U_1, V_1, M_1, Z_1)$ and
eigenvalue $\omega$. Since this is a linear problem, we are allowed to
take the amplitude of one variable arbitrarily, for which we fix ${\rm
Re}(a_0)={\rm Im}(a_0)=1$ at the magnetosonic point.  We choose three
trial complex values for $\omega$, $b_0$, and $d_0$. This specifies the
values of $S_1$, $U_1$, $V_1$, and $M_1$ as well as their derivatives
at $x=\xSO$ from Equation \eqref{e:exp}.  We then integrate Equations
\eqref{e:ds1}--\eqref{e:db1} from $x=\xSO$ both in the downstream
direction up to $x=\xSH+L$ and in the upstream direction to $x=\xSH$,
and apply the periodic conditions to the perturbation variables. At the
shock front, the fourth boundary condition (Eq.~\eqref{e:j4})
determines $Z_1$, which is in turn used to check the other three
boundary conditions. If Equations \eqref{e:j1}--\eqref{e:j3} are not
satisfied within tolerance, we return to the first step and repeat the
calculations by changing $b_0$, $d_0$, and $\omega$, one by one, until
all the perturbed jump conditions are met.

\section{Results}\label{sec:disp}

\subsection{One-dimensional Modes with $\ky=0$}\label{sec:onedim}

We first apply the method described above to 1D perturbations with
$\ky=0$. Tables \ref{t:eigF05} and \ref{t:eigF10} list the ten lowest
eigenfrequencies with differing $\BBeta$ for $\F=5\%$ and $\F=10\%$,
respectively. The modes are numbered in the ascending order of $\omgR$.
Strongly magnetized shocks with small $\BBeta$ possess a small number
of modes. In most cases, spiral shocks have a pure decaying mode (with
$\omgR=0$ and $\omgI<0$), a single overstable mode (with $\omgR\neq0$
and $\omgI>0$), and multiple underdamping modes (with $\omgI<0$). The
behavior of the growth rate of the overstable mode with $\BBeta$ is not
simple. In models with $\F=5\%$, magnetic fields tend to stabilize the
1D overstable mode, suppressing it completely  when $\BBeta\leq 3$. In
models with $\F=10\%$, on the other hand, $\omgI$ of the overstable
mode is almost independent of $\BBeta$. Figure \ref{f:1eigfun} compares
the eigenfunction $S_1$ between $\BBeta=\infty$ and $10$ cases for five
odd-$n$ modes, showing that magnetic fields do not much affect the
shapes of the eigenfunctions. The number of nodes in $S_1$ is $2(n-3)$
for $n\ge 5$ regardless of $\beta$.

To verify the growth rates of the overstable modes found by our
stability analysis, we run direct MHD simulations by utilizing the
Athena code \citep{sto08,sto09}.  Among various schemes implemented in
Athena, we use the constrained corner transport method for
directionally unsplit integration, the HLLE Riemann solver for flux
computation \citep{har83,ein91}, and the piecewise linear method
for spatial reconstruction. The simulation domain is a 1D box with
length $L$ resolved by 2048 zones. We start from an initially uniform
surface density $\Sigma_c$ and uniform magnetic fields with $\BBeta=1$
in the local frame described in Section \ref{sec:steady}. We slowly
turn on the spiral arm potential amplitude and make it attain full
strength $\F=10\%$ at $t=40/\Omega$.

Figure \ref{f:1dsim}(a) plots the temporal evolution of the gas surface
density at $x/L=0$. Note that $\Sigma$ increases with time as the arm
strength increases and saturates at about $t=50/\Omega$, after which
$\Sigma$ exhibits a driven-oscillator behavior. The red solid lines
enveloping the density fluctuations after saturation correspond to the
growth rate of $1.1\times10^{-2}\Omega$, very close to the value given
in Table \ref{t:eigF10}. The inset clearly shows the oscillations of
$\Sigma$ over the time interval $220 \leq t\Omega \leq 260$. Figure
\ref{f:1dsim}(b) plots the Fourier-transformed power spectrum over
$t\Omega=200$--$400$. It is peaked at some specific frequencies,
indicated by red arrows, equal to the real parts of the
eigenfrequencies listed in Table \ref{t:eigF10}. Note the dominance of
the $n=3$ mode with $\omgR/\Omega=2.88$ in the power spectrum.  This
confirms that the results of our linear stability analysis are reliable
at least for the 1D perturbations.

It is uncertain what causes the 1D overstability of spiral shocks. It
appears that the overstability arises due to a complicated interplay
among various involved agents such as spiral forcing, epicycle motions,
thermal pressure, magnetic fields, non-uniform background density and
shear, etc. When a spiral shock is displaced from its equilibrium
position, it is forced to move backward, but with an increased
amplitude. Notwithstanding its nature, the growth time of the
overstability amounts to $\sim 4 \Gyr \;(\Omega/26\freq)^{-1}$, which
is much longer than that of the 2D wiggle instability presented below.
This is also longer than the expected lifetime (shorter than $\sim 1
\Gyr$) of spiral arms (e.g., \citealt{sel84,oh08,oh15,spe11,spe12}).
Therefore, spiral shocks can be considered stable to 1D perturbations
for all practical purposes.

\subsection{Two-dimensional Modes $\ky\neq0$}\label{sec:twodim}

\subsubsection{Dispersion Relations}\label{sec:twodisp}

We now consider 2D perturbations with $\ky\neq0$ to explore the WI of
magnetized spiral shocks. Figure \ref{f:disp2d} plots the dispersion
relations of ten lowest-frequency eigenmodes over $0\leq \ky L\leq60$
for a spiral shock with $\F=5\%$ and $\BBeta=10$. Similarly to the 1D
modes, these 2D modes are numbered in the ascending order of $\omgR$ at
$\ky=0$. The solid and dashed lines correspond to $\omgI$ and $\omgR$,
respectively. Note that $\omgR$ varies almost linearly with $\ky$,
which indicates that the eigenmodes can be expressed as a linear
superposition of entropy-vortex waves and MHD waves (Paper~I). Note
also that each mode becomes unstable (i.e., $\omgI>0$) in a few ranges
of $\ky$, although the corresponding growth rates for all $n$ are less
than $0.5\Omega$. This is in contrast to the unmagnetized cases in
which $\omgI$ of the $n=7$ mode keeps increasing with $\ky$ (see Fig.\
4 of Paper~I).

To evaluate the quantitative effects of the magnetic fields on the WI,
we compare the dispersion relations of overstable modes with differing
$\BBeta$ for the $\F=5\%$ and $\F=10\%$ cases in Figures
\ref{f:disp2dF05} and \ref{f:disp2dF10}, respectively. The growth rates
and wavelengths of the most unstable modes depend on $\F$ as well as
$\BBeta$ quite sensitively. The WI of unmagnetized arms is dominated by
a single ($n=7$ for $\F=5\%$ and $n=4$ for $\F=10\%$) mode, and this
holds as long as magnetic fields are relatively weak with
$\BBeta \geq 100$ for $\F=5\%$ and $\BBeta\geq 5$ for $\F=10\%$.
Clearly, magnetic fields reduce both the maximum growth rate
$\omgI_{\rm max}$ and the corresponding wavenumber $k_{y, \rm max}$,
such that $\omgI_{\rm max}/\Omega= 1.36$ and $1.13$ occurring at $k_{y,
\rm max} L=100.6$ and $78.1$ for $\BBeta=\infty$ and $100$,
respectively, when $\F=5\%$. For $\F=10\%$, these values are increased
to $\omgI_{\rm max}/\Omega= 4.71$, 2.53, 1.46, and 0.90 occurring at
$k_{y,\rm max} L=198.3$, 90.5, 47.1, and 31.9 for $\BBeta=\infty$, 100,
10, and 5, respectively. For more strongly magnetized arms with
$\BBeta\leq 10$ for $\F=5\%$ and $\BBeta\leq 3$ for $\F=10\%$, on the
other hand, there is no single dominant mode, but spiral shocks are
unstable to several different modes with similar growth rates that
become smaller with decreasing $\BBeta$ and $\F$. In this case, the
range of the most unstable wavenumbers is $k_{y, \rm max}L\sim20$--$50$
for $\F=5\%$ and $k_{y, \rm max}L\sim5$--$30$ for $\F=10\%$, largely
independent of $\BBeta$. For arms with $\BBeta=1$, $\omgI_{\rm
max}/\Omega \leq 0.25$ and $0.20$ for $\F=5\%$ and $10\%$,
respectively. The reduction of the growth rates is larger at larger
$\ky$, suggesting that magnetic fields suppress the growth of the WI,
especially for small-scale modes.

Figure \ref{f:2deigen} plots the eigenfunctions $S_1$, $U_1$, $V_1$,
$M_1$, and $\Xi_1$ of the most unstable modes with $\omega/\Omega=84.4
+1.13i$ and $\ky L=78.1$ for $\BBeta=100$ in the left panels, and with
$\omega/\Omega=59.1+0.50i$ and $\ky L=53.4$ for $\BBeta=10$ in the
right panels. Also plotted as the black solid lines in the bottom
panels are the solutions of Equation \eqref{e:intpv}, which are in good
agreement with the direct numerical results. The spiral forcing is
fixed to $\F=5\%$ for both cases. When $\BBeta=100$, the amplitudes of
the eigenfunctions decrease monotonically with the distance downstream
from the shock front, similarly to the unmagnetized cases (Paper~I).
For more strongly magnetized spiral shocks, however, the eigenfunctions
do not decay monotonically. Although $|\Xi_1|$ decreases with $x$ right
after the shock front, the magnetic tension and pressure forces from
the perturbed fields cause the fluctuations in the amplitude of $\Xi_1$
further downstream.

Figure \ref{f:2deigenmap} overlays the configuration of the
perturbed magnetic fields shown as black solid lines over the real
perturbed PV constructed as
 \be
   {\rm Re}(\xi_1) =
         {\rm Re}(\Xi_1)\cos (\ky y ) - {\rm Im}(\Xi_1)\sin (\ky y),
 \ee
in the regions with $-0.2\leq  x/L \leq 0.3$ and $0\leq y/\lambda_y
\leq 1$, with $\lambda_y=2\pi/\ky$, for the models shown in Figure
\ref{f:2deigen}.\footnote{Note that the configurations of the
perturbed field lines shown in Figure \ref{f:2deigenmap} are only for
an illustrative purpose: they do not well trace the total field lines
when the unperturbed component is stronger than the perturbed one.}
White dots in both panels trace the wavefronts of the perturbed PV,
clearly showing a discontinuity in $\kx$ across the shock located at
$x/L=-0.07$ and $-0.08$ for $\BBeta=100$ and 10, respectively. Gas
motions associated with entropy-vortex modes and MHD modes not only
bend magnetic fields but also compress them in the postshock flows.
This in turn breaks the conservation of the PV and results in a
non-monotonic behavior of $|\Xi_1|$. Consequently, the net reduction of
the perturbed PV from one shock to next is smaller for shocks with
stronger magnetic fields, which should be matched by the jump of the
perturbed PV, $\jump{\Xi_1}$, across the shock front. Note that the
perturbed fields are strong inside the regions of positive ${\rm
Re}(\xi_1)$ created by the WI, and reverse the directions in the
regions between them.

\subsubsection{Effects of Magnetic Fields on the WI}\label{sec:physical}

For unmagnetized shocks, Paper~I derived an analytic expression for the
change of the perturbed PV across a disturbed shock, demonstrating that
the shock distortion is indeed a source of the PV (see also
\citealt{hay57,kel97}). Appendix \ref{a:jumpPV} derives a similar
expression by combining Equations \eqref{e:j1}--\eqref{e:j3} for the
perturbed PV jump across a magnetized spiral shock. The derived
Equation \eqref{e:ppvj2} recovers Equation (A8) of Paper~I when
$\BBeta=\infty$.

While $\jump{\Xi_1}$ formally depends on all of the five perturbation
variables, we find empirically that the $U_1$, $V_1$, and $Z_1$ terms
dominate in Equation \eqref{e:ppvj2}. Since $|d\uT/dx| \ll |\omgD^s|$
for overstable modes, one can approximately write
 \be\label{e:ppvj4}
 \frac{\jump{\Xi_1}} { i\ky}
    \approx    (\mQ_{U, \rm H} + \mQ_{U, \rm M})  \frac{U_1^\spre}{\Sigma_0^\spre}
              + \mQ_{V, \rm H} \frac{V_1^\spre}{\Sigma_0^\spre}
              +(\mQ_{Z, \rm H} + \mQ_{Z, \rm M}) \frac{Z_1^\spre}{\Sigma_0^\spre},
 \ee
where the coefficients are given as
 \be\label{e:Ucoeff1}
  \mQ_{U, \rm H} = \frac{(\mu-1)^2}{\mu^2},
 \ee
 \be\label{e:Ucoeff2}
  \mQ_{U, \rm M} =  2\mathcal{A}\mathcal{B}
  \left(1+\frac{1}{\mu^2} + \frac{(\mu+1)\mathcal{B}}{\mu}\right),
 \ee
 \be\label{e:Vcoeff}
  \mQ_{V, \rm H} = - \frac{q_c\Omega L}{\uT^\spre},
 \ee
 \be\label{e:Zcoeff1}
  \mQ_{Z, \rm H} = i\omgD^s \frac{(\mu-1)^2}{\mu^2},
 \ee
 \be\label{e:Zcoeff2}
  \mQ_{Z, \rm M} = \frac{2i\omgD^s\mathcal{A}\mathcal{B}}{\mu(\mu+1)}
  \left[ (\mu+1)^2 - \mu(\mu-1)\mathcal{B} \right],
 \ee
with $\mathcal{A}$ and $\mathcal{B}$ defined by Equations \eqref{e:AA}
and \eqref{e:BB}, respectively.  The subscripts ``H'' and ``M'' in the
coefficients stand for the hydrodynamic and magnetic contributions to
the perturbed PV, respectively. For unmagnetized shocks,
$\mathcal{B}=0$, so that $\mQ_{U, \rm M}=\mQ_{Z,\rm M}=0$.

Note that the $U_1$ term in Equation \eqref{e:ppvj4} results from the
tangential variation of the velocity perpendicular to the shock, which
plays a stabilizing role by reducing the PV by shock compression. The
$\mQ_{U, \rm M}$ term explains the PV change by the perturbed magnetic
pressure (Eq.~\eqref{e:rk2}). That both $\mQ_{U, \rm H}$ and $\mQ_{U,
\rm M}$ are always positive implies that the perturbed magnetic
pressure stabilizes the WI. This can be seen more explicitly by
considering a special case where the PV is retained only in $U_1$, with
the other perturbation variables taken zero. Then, one can show that
$\Xi_1^\spost/\Xi_1^\spre = (2\mu-1)/\mu^2 - \mQ_{U, \rm M}$, which is
always less than unity. The $V_1$ term in Equation \eqref{e:ppvj4}
comes simply from the discontinuity of the radial wavenumber across the
shock, independent of the magnetic fields, which also stabilizes the WI
(Paper~I).

On the other hand, the $Z_1$ term originates from the shock deformation
along the tangential direction. This is a source for the PV generation,
leading to the WI. The $\mQ_{Z, \rm M}$ term is due to the perturbed
magnetic stress (Eq.~\eqref{e:rk3}). Since $\mQ_{Z, \rm H}$ and
$\mQ_{Z, \rm M}$ always have the same sign, the stress of the deformed
magnetic fields destabilizes the WI. Therefore, the role of the
perturbed magnetic fields differs in the $U_1$ and $Z_1$ terms. As
plotted in Figure \ref{f:coeff}, however, the ratios $\mQ_{U, \rm
M}/\mQ_{U, \rm M}$ and $\mQ_{Z, \rm M}/\mQ_{Z, \rm H}$ are quite small,
especially for large $\BBeta$ and $\F$.  The maximum contribution of
the magnetic terms is less than 30\% of the hydrodynamic terms, which
occurs when $\F=5\%$ and $\BBeta=1$. This indicates that the effects of
\emph{perturbed} magnetic fields themselves to the WI are not
significant.

What is then responsible for the reduction of the growth rates in the
presence of magnetic fields, as  shown in Figures \ref{f:disp2dF05} and
\ref{f:disp2dF10}? It is through the magnetic pressure of the
\emph{unperturbed} background fields that tend to reduce the
compression factor $\mu$ (Table \ref{t:1d}; see also Figure 3 of
\citet{lee14}). The amount of the PV production is smaller when a
shock is weaker for fixed $Z_1$ (Eqs.~\eqref{e:Zcoeff1} and
\eqref{e:Zcoeff2}). More importantly, the amplitudes of the
perturbation variables obtained by integrating Equations
\eqref{e:ds1}--\eqref{e:db1} decay less with $x$ for smaller $\mu$
(e.g., Fig.~\ref{f:2deigen}), resulting in larger values of
$|U_1^\spre|$ and $|V_1^\spre|$ in more strongly magnetized shocks.
This enhances the stabilizing role of the velocity terms relative to
the destabilizing $Z_1$ term in Equation \eqref{e:ppvj4}. Consequently,
spiral shocks with stronger magnetic fields become more stable to the
WI.

\subsection{Numerical Simulation}\label{sec:num}

To check the results of our 2D linear stability analysis, we run direct
MHD simulations using the Athena code. As a simulation domain, we set
up a rectangular box with size $L\times 2L$ that is resolved by a
$2048\times4096$ grid.\footnote{By also running models with size
$L\times L$, we have confirmed that the numerical results are
insensitive to the domain size along the $y$-direction since $k_{y, \rm
max}L \ll 1$.} Initially, we construct a 1D shock profile found in
Section \ref{sec:equil} as a background state. We then apply
small-amplitude perturbations to the background density that are
realized by a flat-power Gaussian random field with a standard
deviation of $10^{-3}\Sigma_0$. We take the shearing-periodic boundary
conditions at the $x$-boundaries and the periodic boundary conditions
at the $y$-boundaries (e.g, \citealt{haw95,kim02a,kim06}).

Figure \ref{f:2dsim} displays snapshots of the gas surface density in
logarithmic scale and the configurations of the magnetic fields in the
regions with $-0.2\leq x/L \leq 0.35$ and $0.5\leq y/L\leq 1.2$ for
models with $\BBeta=100$, $10$, $5$, and $3$ at $t\Omega=8$, $16$,
$25$, $40$ when the WI saturates, respectively. The number of nonlinear
structures along the $y$-direction that develop most strongly over the
simulations domain is $21$, $15$, $12$, and $9$, corresponding to the
wavenumber of $k_{y, \rm max}L=66.0$, $47.1$, $37.7$, and $28.3$ for
$\BBeta=100$, $10$, $5$, and $3$, respectively. Note that the magnetic
fields bend around nonlinear structures that are more strongly
magnetized than the surrounding regions.

Figure \ref{f:growth} compares the temporal evolution of the maximum
surface density measured at $x/L=0$ in these models. The fastest
growing mode in each model has a slope of $0.48$, $0.22$, $0.17$, and
$0.13$, corresponding to the growth rate of $\omgI_{\rm
max}/\Omega=1.11$, $0.50$, $0.39$, and $0.30$, for $\BBeta=100$, $10$,
$5$, and $3$, respectively. These numerically-measured wavelengths and
growth rates are marked as star symbols in Figure \ref{f:disp2dF05}, in
good agreement with the analytic results for the $n=7$, 10, 9, and 7
modes, respectively. Since various overstable modes with different $n$
have similar maximum growth rates, which mode the system picks up
should also depend on the initial power imposed by specific
perturbations. In the model with $\BBeta=3$, for example, the initial
density perturbations with $\ky L=25.1$ corresponding to the maximum
growth rate of the $n=7$ mode are about an order of magnitude larger
than those with $\ky L=37.8$ corresponding to the peak of the $n=9$
mode, emerging most strongly in the nonlinear stage, despite having a
slightly ($\sim6.7\%$) smaller growth rate.

Finally, we remark on the level of turbulence generated by the WI. In
the simulations described above, the density-weighted velocity
dispersion, $\sigma_y = (\int \Sigma ( v - v_0 )^2 dxdy/\int \Sigma
dxdy)^{1/2}$, in the direction parallel to the arms is found to be
$\sim 1.4, 1.1, 0.8,$ and $0.7 \kms$ at the time when the WI saturates
for $\BBeta=100$, $10$, $5$, and $3$, respectively, which are
interestingly equal approximately to $\omgI_{\rm max}/k_{y,\rm
max}$.\footnote{The density-weighted velocity dispersion in the
direction perpendicular to the arm is not entirely due to the WI
because of the contamination by 1D overstable modes that make the shock
move back and forth in the $x$-direction (Section \ref{sec:onedim}).}
Due to nonlinear interactions and mergers of clumps created by the WI,
$\sigma_y$ increases further by about a factor of 2. This suggests that
the WI can be a considerable, although not dominant, source of
turbulence energy in the ISM since the energy injection occurs to the
densest part of the gas.

\section{Summary and Discussion}\label{sec:summary}

We have investigated the WI of magnetized spiral shocks in a galactic
disk using both a linear stability analysis and direct MHD simulations.
This is a straightforward extension of Paper~I that studied the case of
unmagnetized spiral shocks. We assume that the gas disk is
infinitesimally thin, isothermal, and non-self-gravitating. We
parameterize the strengths of the stellar spiral arms and magnetic
fields using the dimensionless parameters $\F$ and $\BBeta$,
respectively (Eqs.~\eqref{e:F} and \eqref{e:betac}). As background
states, we first obtain the steady spiral shock profiles with differing
$\F$ and $\BBeta$. We then impose small-amplitude wiggling
perturbations to the equilibrium spiral shocks, and calculate their
dispersion relations as a boundary-value problem with eigenfrequencies.
Our local MHD simulations readily pick up the most unstable modes for a
given set up parameters, with the numerical growth rates very close to
the results of the linear stability analysis.

The existence of the overstable modes proves that the WI is physical in
origin, resulting from the accumulation of the perturbed PV from a
distorted shock front. Our results show that magnetic fields suppress
the WI, but not completely, at least for $\BBeta\geq1$.  The
stabilizing role of magnetic fields is not from the perturbed fields
but directly from the background unperturbed fields that tend to reduce
the shock compression factor $\mu$ by exerting magnetic pressure. When
magnetic fields are relatively weak, the WI is dominated by a single
dominant mode. When $\F=5\%$, the most unstable mode has a growth rate
$\omgI_{\rm max}/\Omega= 1.36$ and 1.13 occurring at $k_{y, \rm
max}L=100.6$ and 78.1 for $\BBeta =\infty$ and 100, respectively. When
$\F=10\%$, these values are increased to $\omgI_{\rm max}/\Omega=4.71$,
2.53, 1.46, and 0.90 occurring at $k_{y, \rm max} L=198.3$, 90.5, 47.1,
and 31.9 for $\BBeta =\infty$, 100, 10, and 5, respectively. For more
strongly magnetized arms with $\BBeta\leq 10$ for $\F=5\%$ and
$\BBeta\leq 3$ for $\F=10\%$, on the other hand, several overstable
modes have similar growth rates that become smaller with decreasing
$\BBeta$, while the wavelength range of the most unstable modes is
$k_{y, \rm max}L \sim 20$--50 for $\F = 5\%$ and $k_{y, \rm max}L \sim
5$--30 for $\F = 10\%$, insensitive to $\BBeta$.

We have found that magnetic fields stabilize the WI at small scales by
reducing the shock compression factor. When $\BBeta=10$, the most
unstable wavelength is decreased by a factor of about 2 and 4 for
$\F=5\%$ and 10\%, respectively, compared to the unmagnetized cases.
The stabilization of the WI by magnetic fields has already been
observed in previous nonlinear simulations of spiral galaxies. For
instance, \citet{kim06} ran 2D local models for the formation of
feathers and provided the magnetic field topology at the nonlinear
stage resulting from the vorticity generation near the shock front.
They further showed that vortical clumps produced merge into massive
clouds with mass $\sim10^7\Msun$ each in the presence of self-gravity.
Grid-based global simulations by \citet{she06} found no indication of
the development of the WI in a model with $\BBeta=1$ and $\F=10\%$.
Based on our results, this is presumably not because magnetic fields
completely suppress the WI but because its growth time is longer than
the simulation time span (two orbits) in their models with very
weak ($\sim0.1\%$) initial perturbations.\footnote{Figure
\ref{f:disp2dF10} indicates that the maximum growth rate of the WI is
$\omgI_{\rm max}\simeq 0.25\Omega$ for $\BBeta=1$ and $\F=10\%$. Thus,
the amplitude of the perturbations that start out initially at $0.1\%$
is expected to grow via WI to $10^{-3}\exp(0.25\times4\pi) \sim 0.02$
after two orbits, which is too small to be evident in the models of
\citet{she06}.} By running particle-based simulations, on the other
hand, \citet{dob08} found non-axisymmetric structures, albeit weak,
sill grow in models with $\BBeta=1$. This is probably because they
might have large-amplitude density perturbations arising from the
Poisson noises in the initial particle distributions, helping the WI
readily manifest in their simulations.

It is interesting to compare the predicted wavelength $\lambda_{y, \rm
max}=2\pi/k_{y, \rm max}$ from our linear stability analysis with the
observed spacings of feathers. By analyzing the \emph{Hubble} archival
data, \citet{lav06} measured the feather spacings in grand-design
spiral galaxies. They reported that M51 and M74 have the feather
spacing that increases from $\sim 0.2\kpc$ in the inner regions ($R\sim
1\kpc$) to $\sim 1\kpc$ in the outer regions ($R\sim10\kpc$). Adopting
the arm pitch angles of $i=21.1^\circ$ for M51 \citep{she07} and
$i=15.7^\circ$ for M74 \citep{gus13} (see also \citealt{hon15}), these
correspond to $\lambda_{y, \rm max}/L \sim0.1$--$0.2$, consistent with
our results for magnetized arms with $\BBeta \lesssim 10$ for $\F=5\%$
or with $\BBeta \sim 5$--$10$ for $\F=10\%$.\footnote{In M51, the
total field strength in the arm regions is $\sim20$--$25\mu$G
\citep{fle11}. Taking the mean gas surface density of
$10^2$--$10^{2.5}\Msun\pc^{-2}$ \citep{mei13} and a disk thickness of
$H=200\pc$, this corresponds to $\beta_{\rm arm}\sim 0.4$--$1.7$.}
\citet{elm83} presented the separations of \ion{H}{2} regions and
\ion{H}{1} superclouds along arms in various spiral galaxies. Taking
$m=2$ and $i=20^\circ$ arbitrarily, the observed separations listed in
their Table 2 are distributed in the range $\lambda_{y, \rm
max}/L=0.1$--$1$, with a mean value of $\sim0.4$, a factor of $\sim
2$--$4$ larger than the mean feather spacings mentioned above. This may
indicate that \ion{H}{2} regions and \ion{H}{1} superclouds represent
highly nonlinear structures created by mergers of WI-induced feathers.

Recently, \citet{pue14} introduced a new method to \emph{locally}
determine the distributions of the pitch angles of spiral arms and
their substructures. Applying the method to 8 $\mu$m \emph{Spitzer}
images of M51 and M81, they found that the mean difference between the
pitch angles of the main spiral arms and the interarm secondary
structures is $\Delta i \sim 10^\circ$--$30^\circ$. We apply the same
method to the results of our MHD simulations to calculate the pitch
angles of the nonlinear structures resulting from the WI off the main
arms. Figure \ref{f:pitch} plots $\Delta i$ when the structures
saturate in models with $\BBeta=3$--100 and $\F=5\%$, averaged over the
immediate postshock regions from $x=\xSH$ to $x=\xSH+L/4$. Note that
$\Delta i$ is smaller in more strongly magnetized models. This is
mostly because of shear reversal in the postshock regions, discussed in
Section \ref{sec:equil}, which is larger for stronger shocks. Since
shear reversal tends to decrease $\kx$, $\Delta i = \tan^{-1}(\ky/\kx)$
is smaller for weaker shocks with smaller $\BBeta$. The numerical
values of $\Delta i \sim 20^\circ$--$35^\circ$ for $\BBeta\sim 3$--$10$
are close to the values obtained by \citet{pue14} for M51 and M81.

As discussed above, the theoretical predictions of our analysis based
on 2D, non-self-gravitating models are similar to the observed
properties of feathers, suggesting that the WI may be responsible for
the feather formation in the presence of magnetic fields. However, it
still remains to be seen whether the WI would grow into observed
interarm features in real disk galaxies for the following reasons.
First, \citet{kim06} showed that spiral shocks in vertically-stratified
disk exhibit non-steady flapping motions in the direction perpendicular
to the arms. Together with strong vertical shear, these shock flapping
motions tend to disrupt coherent vortical structures and thus prevent
the growth of the WI.

Second, \citet{lee12} and \citet{lee14} recently conducted a parameter
study of feathering instability of a magnetized, self-gravitating
spiral shock. In particular, \citet{lee14} found that the most
unstable mode of the feathering instability is $\sim530\pc$ in their
M51 arm, similar to the observed feather spacing reported by
\citet{lav06}. He also found that the growth rate of the
feathering instability is typically $\sim\Omega$, which is larger than
the growth rate ($\sim 0.2$--$0.3\Omega$) of the WI for $\beta=1$
studied in this work. Without considering the effects of the
background shear, however, they were unable to study the combined
effects of the wiggle and feathering instabilities. It is possible that
the WI grows fast and provides seeds for the onset of the feathering
instability. Or, the WI is completely suppressed by shock flapping
motions in stratified disks. To fully understand the feather
formation, therefore, it is necessary to study the stability of spiral
shocks in a vertically extended, self-gravitating, magnetized disk that
harbor not only the wiggle and feathering instabilities but also the
Parker instability.

\acknowledgments

We acknowledge helpful discussions with J.~Kim and D.~Ryu. We are also
grateful to the referee for a thoughtful report. This work was
supported by the National Research Foundation of Korea (NRF) grant, No.
2008-0060544, funded by the Korea government (MSIP). The computation of
this work was supported by the Supercomputing Center/Korea Institute of
Science and Technology Information with supercomputing resources
including technical support (KSC-2014-C3-003).

\appendix

\section{Potential Vorticity}\label{a:vor}

Here, we derive an equation for the evolution of the perturbed PV.
Equation \eqref{e:mom} can be decomposed as
 \be\label{e:decom1}
  \frac{Du_T}{Dt}=-\cs^2\frac{\partial\ln\Sigma}{\partial x}
  +2\Omega v-\frac{d\Phi_s}{dx}-\frac{B_y}{4\pi\Sigma}
  \left(\frac{\partial B_y}{\partial x}-\frac{\partial B_x}{\partial y}\right),
 \ee
 \be\label{e:decom2}
  \frac{Dv_T}{Dt}=-\cs^2\frac{\partial\ln\Sigma}{\partial y}
  -\frac{\kappa^2}{2\Omega}-q_c\Omega u_T+\frac{B_x}{4\pi\Sigma}
  \left(\frac{\partial B_y}{\partial x}-\frac{\partial B_x}{\partial y}\right),
 \ee
where
  \be
    \frac{D}{Dt} = \frac{\partial}{\partial t} + \vf_T\cdot\nabla
  \ee
is the Lagrangian time-derivative, and we have used the identity
$\vf_T\cdot\nabla \vf=\vf_T\cdot\nabla\vf_T+q_c\Omega u_T$.

Differentiating Equations \eqref{e:decom1} and \eqref{e:decom2} with
respect to $y$ and $x$, respectively, and then subtracting the
resulting equations, we obtain
 \be\label{e:pvres1}
  \frac{D}{Dt}\left(\nabla\times\vf_T+2\Omega\right) =
  \left(\nabla\times\vf_T+2\Omega\right)\frac{D}{Dt}\ln\Sigma
  +\frac{\mathbf{B}\cdot\nabla}{4\pi}
  \left(\frac{\nabla\times\mathbf{B}}{\Sigma}\right),
 \ee
which can be simplified to
 \be\label{e:pvres2}
  \frac{D \boldsymbol{\xi}}{Dt} = \frac{\mathbf{B}\cdot\nabla}{4\pi\Sigma}
  \left(\frac{\nabla\times\mathbf{B}}{\Sigma}\right),
 \ee
where $\boldsymbol{\xi}$ is the PV defined in Equation \eqref{e:pv}.
Equation \eqref{e:pvres2} states that $\boldsymbol{\xi}$ is in general
not conserved in magnetized flows, which is in contrast to the
hydrodynamic case where $D\boldsymbol{\xi}/Dt=0$.

After linearizing Equation \eqref{e:pv}, one can write the perturbed PV
as
 \be
  \xi_1=\frac{|\nabla\times\mathbf{v_1}|}{\Sigma_0}
  -\xi_0\frac{\Sigma_1}{\Sigma_0},
 \ee
where $\xi_0$ is the PV in the background unperturbed shock
(Eq.~\eqref{e:pv0}). Analogous to Equation \eqref{e:ptb}, we define the
amplitude $\Xi_1$ of $\xi_1$ as
 \be\label{e:pv1}
  \Xi_1  \equiv \xi_1 (x,y,t) e^{i\omega t - i\ky y}
  =\frac{1}{\Sigma_0}\left(\frac{dV_1}{dx}-i\ky U_1\right)-\xi_0 S_1.
 \ee
In terms of the perturbation variables, Equation \eqref{e:pvres2} then
becomes
 \be\label{e:pvcon1}
  \left(-i\omgD + \uT\frac{d}{dx}\right)\Xi_1=
  \frac{i\ky\cs^2}{\BBeta\Sigma_0}
  \left[
    \left(\frac{d^2}{dx^2}\ln\Sigma_0 +\ky^2 \right) M_1
  - \left(\frac{1}{\Sigma_c}\frac{d\Sigma_0}{dx} S_1 + \frac{d^2 M_1}{dx^2}\right)
    \right].
 \ee
The terms in the first parentheses in the right-hand side of Equation
\eqref{e:pvcon1} are due to magnetic tension, while those from magnetic
pressure are given in the second parentheses.

A formal solution of Equation \eqref{e:pvcon1} is given by
 \be\label{e:intpv}
  \Xi_1(x)=\,\,(\Xi_1^\spost+\Psi_{\rm T}+\Psi_{\rm P})\cdot
  e^{-\tau\omgI} \exp \left(i\int_{\xSH}^{x} k_{x,v}(x)dx\right)
 \ee
where $\Xi_1^\spost$ is the PV at the immediate postshock regions,
$\tau = \int_{\xSH}^{x} dx/\uT$ is the Lagrangian time elapsed from the
shock front, and $\kx (x) = (\omgR-\vT\ky)/\uT$ is the local
$x$-wavenumber. In Equation \eqref{e:intpv}, $\Psi_{\rm T}$ and
$\Psi_{\rm P}$ are defined by
 \be\label{e:PsiT}
  \Psi_{\rm T} \equiv
  \frac{i\ky\cs^2}{\BBeta\uT\Sigma_0}
  \int_{\xSH}^{x}
  \left(\frac{d^2}{dx'^2}\ln\Sigma_0 +\ky^2 \right) M_1
   \cdot e^{\tau(x')\omgI} \exp
  \left(-i\int_{\xSH}^{x'} \kx (x'')dx''\right)dx',
 \ee
and
 \be\label{e:PsiP}
  \Psi_{\rm P}\equiv
  - \frac{i\ky\cs^2}{\BBeta\uT\Sigma_0}
  \int_{\xSH}^{x}
  \left(\frac{1}{\Sigma_c}\frac{d\Sigma_0}{dx'} S_1 + \frac{d^2 M_1}{dx'^2}\right)
  \cdot e^{\tau(x')\omgI}
  {\rm exp}\left(-i\int_{\xSH}^{x'} \kx (x'')dx''\right)dx',
 \ee
representing the contributions to the perturbed PV due to magnetic
tension and magnetic pressure, respectively.

\section{Method of obtaining 1D Equilibrium Shock Profile}\label{a:1d}

The coefficient of $d\uT/dx$ in the left-hand side of Equation
\eqref{e:momx01} vanishes at the magnetosonic point, $x=\xSO$, where
 \be\label{e:singular}
  \left(\uT-\frac{\cs^2}{\uT}-\frac{\cs^2u_c}{\uT^2\BBeta}\right)_{x=\xSO} =0,
 \ee
which is a cubic equation in $\uT$. A subsonic gas should accelerate to
a supersonic speed through the magnetosonic point. To obtain smooth
solutions, we expand $\uT$ and $v_0$ around $x=\xSO$ as
\begin{mathletters}\label{e:so0}
 \bea
  \uT&=&\cA+\alpha_1\DX+\alpha_2\DX^2+\mathcal{O}(\DX^3),\label{e:sou0}\\
  v_0&=&\gamma_0+\gamma_1\DX+\gamma_2\DX^2+\mathcal{O}(\DX^3),\label{e:sov0}
 \eea
\end{mathletters}
where $\DX \equiv x-\xSO$ and $\cA$ denotes the positive real root of
Equation \eqref{e:singular} for given $u_c$, $\cs$, and $\BBeta$.  The
coefficients $\alpha_{1,2}$ and $\gamma_{0,1,2}$ are to be determined
by series expansions near $\DX=0$.

Substituting Equation \eqref{e:so0} in Equations \eqref{e:momy0} and
\eqref{e:momx01} and keeping terms up to the second order in $\DX$, one
obtains
 \be\label{e:bet0}
  \gamma_0 = -\frac{\F}{2}R\Omega\sin\left(\frac{2\pi\xSO}{L}\right),
 \ee
 \be\label{e:alp1}
  \alpha_1 = \left(3-\frac{\cs^2}{\cA^2}\right)^{-1/2}
  \left[2\Omega\gamma_1+\frac{2\F\Omega^2}{\sin i}
  \cos\left( \frac{2\pi\xSO}{L}\right)\right]^{1/2},
 \ee
 \be\label{e:bet1}
  \gamma_1=\frac{u_c-\cA}{\cA}\Omega,
 \ee
 \be\label{e:alp2}
  \alpha_2=\frac{1}{3} \left(3-\frac{\cs^2}{\cA^2}\right)^{-1}
  \left[\frac{\alpha_1^2}{\cA}\left(3-2\frac{\cs^2}{\cA^2}\right)
  -\frac{u_c\Omega^2}{\cA^2}+\frac{4\gamma_0}{\alpha_1}
   \frac{\pi^2}{L^2}\Omega \right],
 \ee
and
 \be\label{e:bet2}
  \gamma_2 = -\frac{1+\gamma_1}{2\cA}\alpha_1.
 \ee
With these coefficients, $\uT$ and $v_0$ in Equation \eqref{e:so0} vary
smoothly near the magnetosonic point.

The equilibrium spiral-shock profiles should satisfy the following jump
conditions at the shock front, $x=\xSH$:
\begin{mathletters}\label{e:sj}
\bea
  \jump{\uT\Sigma_0} &=& 0,\label{e:s1} \\
  \jump{(\cs^2+\uT^2)\Sigma_0+\frac{B_0^2}{8\pi}} &=& 0,\label{e:s2} \\
  \jump{v_0} &=& 0, \label{e:s3} \\
  \jump{\uT B_0} &=&0, \label{e:s4}
\eea
\end{mathletters}
where $\jump{f} \equiv f^\spost - f^\spre$, with the superscripts
``$\spost$'' and ``$\spre$'' representing the quantities evaluated at
the immediate behind ($x=\xSH + 0$) and ahead ($x=\xSH -0$) of the
shock front, respectively. Note that Equations \eqref{e:s1} and
\eqref{e:s4} are satisfied automatically from Equations \eqref{e:con0}
and \eqref{e:ind0}.

Equilibrium profiles of magnetized spiral shocks can be constructed as
follows. For given $\F$ and $\BBeta$, we first choose the magnetosonic
point $\xSO$ arbitrarily, and integrate Equations \eqref{e:momy0} and
\eqref{e:momx01} starting from $x=\xSO$ in both the upstream and
downstream directions. We apply the periodic boundary conditions at
$x/L=\pm1/2$ and determine the shock position $\xSH$ by imposing
Equation \eqref{e:s3}. We then check whether Equation \eqref{e:s2} is
also fulfilled. If not within tolerance, we return to the first step to
change $\xSO$ until all the jump conditions are satisfied.

\section{Jump of Potential Vorticity at the Distorted Shock Front}
\label{a:jumpPV}

Here we derive an algebraic expression for the jump condition of the
perturbed PV, $\jump{\Xi_1}=\Xi_1^\spost-\Xi_1^\spre$, at the shock
front ($x=\xSH$) in terms of the preshock variables ($S_1^\spre$,
$U_1^\spre$, $V_1^\spre$, and $M_1^\spre$) and $Z_1$.

With the help of Equation \eqref{e:djump}, the jump conditions
(Eq.~\eqref{e:sj}) of the background steady shock can be combined to
yield
 \be\label{e:aux1}
   \left(\frac{\uT^\spost}{\cs}\right)^2
   = \frac{1}{\mu} + \frac{\mu+1}{2\mu\BBeta}\frac{u_c}{\uT^\spre},
 \ee
for the postshock Mach number, and
 \be\label{e:aux2}
  \mach^2 = \left(\frac{\uT^\spre}{\cs}\right)^2
   =  \mu + \frac{\mu(\mu+1)}{2\BBeta}\frac{u_c}{\uT^\spre},
 \ee
for the preshock Mach number. For later purposes, we calculate
 \be\label{e:app1}
   \frac{d\ln\uT^\spost}{dx} - \frac{d\ln\uT^\spre}{dx}
   = - \mathcal{A} \left[ (\mu +1)^2 +2\mu \mathcal{B} \right]
   \frac{d\ln\uT^\spre}{dx},
 \ee
where
 \be\label{e:AA}
   \mathcal{A} \equiv \left(2\mu + 1 - \frac{\mu^2}{\mach^2}\right)^{-1},
 \ee
and
 \be\label{e:BB}
    \mathcal{B} \equiv 1 - \frac{\mu}{\mach^2}.
 \ee
Note that $\mach=\mu^{1/2}$, $\mathcal{A}=1/(1+\mu)$, and
$\mathcal{B}=0$ for isothermal hydrodynamic shocks. Table \ref{t:1d}
shows that $\mathcal{M}$ is insensitive to $\BBeta$ for given $\F$.
Both $\mathcal{A}$ and $\mathcal{B}$ are positive definite.

It is straightforward to show that Equation \eqref{e:j1} becomes
 \be\label{e:pjs}
   S_1^\spost = S_1^\spre + [U_1^\spre - \mu U_1^\spost - (\mu-1) i
   \omgD^sZ_1]/\uT^\spre,
 \ee
while Equation \eqref{e:j4} is simplified to
  \be\label{e:pjm}
    M_1^\spost = M_1^\spre + (\mu-1) (u_c/\uT^\spre) Z_1.
  \ee
On the other hand, Equation \eqref{e:j2} utilizing Equations
\eqref{e:aux1} and \eqref{e:pjm} becomes
 \be\label{e:pju}
 \begin{split}
   \frac{U_1^\spost}{\mathcal{A}} &=
   - \frac{1}{\mu} \left[ \mu+1 + (\mu+2)\mathcal{B}\right] U_1^\spre \\
    &- \frac{(\mu+1)\mathcal{B}}{\mu}
      \left(\uT^\spre S_1^\spre + \frac{2i\omgD \uT^\spre}{u_c} M_1^\spre\right) \\
       &- i\omgD^2
       \left[ \frac{(\mu+1)^2}{\mu} + (\mu-1)\mathcal{B} \right] Z_1
   +  \left(1+\mu - \mathcal{B} \right)\frac{\mu-1}{\mu} \frac{d\uT^\spre}{dx} Z_1.
 \end{split}
 \ee
Plugging Equation \eqref{e:pju} into Equation \eqref{e:pjs} allows to
write $S_1^\spost$ in terms of the preshock quantities. Lastly,
Equation \eqref{e:j3} together with Equation \eqref{e:pjm} gives
 \be\label{e:pjv}
   V_1^\spost = V_1^\spre - (\mu-1)
   \left[\frac{\kappa^2}{2\Omega} \frac{u_c}{\uT^\spre} -
   i\ky  \left(\frac{\uT^\spre}{\mu} + \frac{u_c}{\BBeta\mach^2} \right)
   \right]  Z_1 + \frac{i\ky(\mu-1)\cs}{\BBeta\mach} M_1^\spre .
 \ee

Combining Equations \eqref{e:dv1} and \eqref{e:pv1}, the jump of the
perturbed PV can be written as
 \be\label{e:ppvj}
 \begin{split}
 \jump{\Xi_1} = & -\frac{i\ky}{\Sigma_0^\spre}
 \left(\frac{U_1^\spost}{\mu} - U_1^\spre \right) -
 \frac{\kappa^2}{2\Omega} \frac{u_c}{(\uT^\spre)^2\Sigma_0^\spre} (\mu
 U_1^\spost - U_1^\spre) \\
 & - \left( \frac{i\ky\uT^\spre}{\mach^2} + \frac{\kappa^2}{2\Omega}
 \frac{u_c}{\uT^\spre} \right) \left(\frac{S_1^\spost}{\Sigma_0^\spre}
 - \frac{S_1^\spre}{\Sigma_0^\spre}\right)
 + \frac{i\omgD^s}{\uT^\spre \Sigma_0^\spre} (V_1^\spost-V_1^\spre) \\
 & -\frac{iq_c\Omega L \ky}{\uT^\spre \Sigma_0^\spre}  V_1^\spre
 - \frac{i\ky}{\BBeta\mach^2 \Sigma_0^\spre}
 \left( \mu\frac{d\uT^\spost}{dx} M_1^\spost - \frac{d\uT^\spre}{dx}
 M_1^\spre\right).
 \end{split}
 \ee
Using Equations \eqref{e:pjs}--\eqref{e:pjv}, we eliminate
$S_1^\spost$, $U_1^\spost$, $V_1^\spost$, and $M_1^\spost$ in Equation
\eqref{e:ppvj} and arrange the terms to obtain
 \be\label{e:ppvj2}
 \frac{ \jump{\Xi_1}} { i\ky }
    = \mQ_S \frac{S_1^\spre}{\Sigma_0^\spre}
              + \mQ_U \frac{U_1^\spre}{\Sigma_0^\spre}
              + \mQ_V \frac{V_1^\spre}{\Sigma_0^\spre}
              + \mQ_M \frac{M_1^\spre}{\Sigma_0^\spre}
              + \mQ_Z \frac{Z_1^\spre}{\Sigma_0^\spre},
 \ee
with the coefficients defined as
 \be\label{e:cqs}
  \mQ_S = \frac{(\mu+1)\mathcal{A}\mathcal{B}}{\mu^2}
  \left(1
  -\frac{\mu^2}{\mach^2}\right) \uT^\spre,
 \ee
 \be\label{e:cqu}
  \mQ_U = \frac{(\mu-1)^2}{\mu^2} +
   2\mathcal{A}\mathcal{B}
  \left(1+\frac{1}{\mu^2} + \frac{(\mu+1)\mathcal{B}}{\mu}\right),
 \ee
 \be\label{cqv}
  \mQ_V=- \frac{q_c\Omega L}{\uT^\spre},
 \ee
 \be\label{cqm}
  \mQ_M=
  \frac{2\mathcal{A}\mathcal{B}}{\mu^2(1+\mu)} \frac{\uT^\spre}{u_c}
  \left\{\left(4\mu^2-1+\frac{2}{\mathcal{A}} \right)\frac{d\uT^\spre}{dx}
  + i\omgD^s \left[1-\mu^2 +
    \mu(1+\mu+2\mu^2)\mathcal{B} \right]\right\},
 \ee
and
  \be\label{cqz}
  \mQ_Z =
  \frac{(\mu-1)^2}{\mu(1+\mu)} \left(1+\frac{1}{\mach}\right)
   \frac{d\uT^\spre}{dx} + i\omgD^s \frac{(\mu-1)^2}{\mu^2}
  + \frac{2i\omgD^s\mathcal{A}\mathcal{B}}{\mu(\mu+1)}
  \left[ (\mu+1)^2 - \mu(\mu-1)\mathcal{B} \right].
 \ee

In the limit of vanishing magnetic fields ($\mach^2\rightarrow\mu$ and
$\mathcal{A}\rightarrow (1+\mu)^{-1}$), $\mQ_S=\mQ_M=0$, and Equation
\eqref{e:ppvj2} reduces to
 \begin{equation}\label{e:ppvj3}
 \frac{\jump{\Xi_1}}{ik_y}  =
  \frac{(\mu-1)^2}{\mu^2} \frac{1}{\Sigma_0^\spre} \left( U_1^\spre
  + \frac{d\uT^\spre}{dx}
  Z_1   + i\omgD^s Z_1 \right)-  \frac{q_c\Omega L}{\uT\Sigma_0^\spre}V_1^\spre,
\end{equation}
identical to Equation (A8) of Paper~I.  Paper~I showed that the
$\omgD^sZ_1$ term in Equation \eqref{e:ppvj3} is responsible for the
production of the PV at a disturbed shock, while the terms involving
$U_1^\spre$ and $V_1^\spre$ suppress the PV due to shock compression
and background shear, respectively.

\begin{deluxetable}{cccccccc}
\tablecaption{Properties of Equilibrium Spiral Shocks\label{t:1d}}
\tablewidth{0pt} %
\tablehead{ \colhead{$\F$} & \colhead{$\beta$} & \colhead{$\xSO/L$}
          & \colhead{$\xSH/L$} & \colhead{$\Sigma_0^\spre/\Sigma_c$}
          & \colhead{$\Sigma_0^\spost/\Sigma_c$}
          & \colhead{$\mu$}
          & \colhead{$\mathcal{M}$}}
\startdata
     &$\infty$& 0.007 & $-0.069$ & $0.546$ & 6.31 & 11.6 & 3.40 \\
     & 100    & 0.008 & $-0.070$ & $0.545$ & 6.12 & 11.2 & 3.41\\
0.05 & 10     & 0.017 & $-0.078$ & $0.538$ & 5.01 & 9.29 & 3.45\\
     & 5      & 0.030 & $-0.085$ & $0.534$ & 4.34 & 8.12 & 3.48 \\
     & 3      & 0.057 & $-0.093$ & $0.528$ & 3.79 & 7.17 & 3.52 \\
     & 1      & 0.193 & $-0.125$ & $0.512$ & 2.62 & 5.12 & 3.63 \\
\hline
     &$\infty$& 0.115 & $-0.005$ & $0.316$ & 10.9 & 34.5 & 5.87 \\
     & 100    & 0.118 & $-0.006$ & $0.316$ & 10.4 & 32.7 & 5.88 \\
0.10 & 10     & 0.143 & $-0.013$ & $0.314$ & 7.80 & 24.8 & 5.92 \\
     & 5      & 0.165 & $-0.019$ & $0.312$ & 6.54 & 20.9 & 5.94 \\
     & 3      & 0.187 & $-0.026$ & $0.311$ & 5.58 & 17.9 & 5.97 \\
     & 1      & 0.249 & $-0.054$ & $0.310$ & 3.70 & 11.9 & 5.99
\enddata
\tablecomments{For the arm and galaxy parameters of $q_c = 1$, $m = 2$,
$\sin i=0.1$, $\Omega_p/\Omega=0.5$, and $\cs/(R\Omega)=0.027$. 
The spiral potential is minimized at $x=0$.}
\end{deluxetable}

\clearpage

\begin{deluxetable}{cccccccccc}
\tablecaption{Eigenfrequencies of One-dimensional Perturbations for
$\F=0.05$\label{t:eigF05}} \tablehead { &
\multicolumn{2}{c}{$\BBeta=\infty$}&
             & \multicolumn{2}{c}{$\BBeta=100$}&
             & \multicolumn{2}{c}{$\BBeta=10$} \\
\cline{2-3}\cline{5-6}\cline{8-9}\\
  mode  & $\omgR/\Omega$& $\omgI/\Omega$ &&
          $\omgR/\Omega$& $\omgI/\Omega$ &&
          $\omgR/\Omega$& $\omgI/\Omega$}
\startdata 1 & 0.000 &$-3.727\times10^{-1}$&& 0.000
&$-3.746\times10^{-1}$&&
    0.000 &$-3.760\times10^{-1}$\\
2 & 0.692 &$-1.782\times10^{-1}$&& 0.695 &$-1.799\times10^{-1}$&&
    0.714 &$-1.859\times10^{-1}$\\
3 & 1.496 &$+1.222\times10^{-3}$&& 1.496 &$+1.123\times10^{-3}$&&
    1.497 &$+5.502\times10^{-4}$\\
4 & 2.628 &$-1.380\times10^{-2}$&& 2.629 &$-1.285\times10^{-2}$&&
    2.640 &$-1.260\times10^{-2}$\\
5 & 4.023 &$-2.758\times10^{-2}$&& 4.037 &$-2.147\times10^{-2}$&&
    4.075 &$-1.942\times10^{-2}$\\
6 & 5.554 &$-2.543\times10^{-2}$&& 5.562 &$-1.935\times10^{-2}$&&
    5.632 &$-1.610\times10^{-2}$\\
7 & 7.144 &$-2.618\times10^{-2}$&& 7.149 &$-1.614\times10^{-2}$&&
    7.249 &$-1.159\times10^{-2}$\\
8 & 8.759 &$-3.077\times10^{-2}$&& 8.767 &$-1.745\times10^{-2}$&&
    8.898 &$-8.692\times10^{-3}$\\
9 &10.389 &$-2.555\times10^{-2}$&&10.406 &$-1.234\times10^{-2}$&&
   10.568 &$-7.080\times10^{-3}$\\
10&12.033 &$-2.250\times10^{-2}$&&12.057 &$-1.056\times10^{-2}$&&
   12.248 &$-6.936\times10^{-3}$\\
\hline\\
 & \multicolumn{2}{c}{$\BBeta=5$} & & \multicolumn{2}{c}{$\BBeta=3$} &
 & \multicolumn{2}{c}{$\BBeta=1$ } \\
\cline{2-3}\cline{5-6}\cline{8-9}\\
  mode  & $\omgR/\Omega$& $\omgI/\Omega$ &&
          $\omgR/\Omega$& $\omgI/\Omega$ &&
          $\omgR/\Omega$& $\omgI/\Omega$\\
\hline\\
1 & 0.000 &$-4.422\times10^{-1}$&& 0.000 &$-6.030\times10^{-1}$&&
    0.965 &$-2.822\times10^{-1}$\\
2 & 0.737 &$-1.920\times10^{-1}$&& 0.772 &$-1.980\times10^{-1}$&&
    1.501 &$-1.595\times10^{-3}$\\
3 & 1.498 &$+1.309\times10^{-4}$&& 1.499 &$-2.745\times10^{-4}$&&
    2.731 &$-8.620\times10^{-3}$\\
4 & 2.652 &$-1.211\times10^{-2}$&& 2.667 &$-1.135\times10^{-2}$&&
    4.363 &$-7.434\times10^{-3}$\\
5 & 4.113 &$-1.803\times10^{-2}$&& 4.160 &$-1.595\times10^{-2}$&&
    6.136 &$-2.720\times10^{-3}$\\
6 & 5.701 &$-1.381\times10^{-2}$&& 5.784 &$-1.103\times10^{-2}$&&-&-\\
7 & 7.348 &$-8.594\times10^{-3}$&& 7.469 &$-4.305\times10^{-3}$&&-&-\\
8 & 9.032 &$-6.711\times10^{-3}$&& 9.185 &$-2.242\times10^{-3}$&&-&-\\
9 &10.727 &$-3.669\times10^{-3}$&&10.920 &$-1.430\times10^{-3}$&&-&-\\
10&12.437 &$-2.614\times10^{-3}$&& - & - &&-&-\\
\enddata
\end{deluxetable}

\begin{deluxetable}{cccccccccc}
\tablecaption{Eigenfrequencies of One-dimensional Perturbations for
$\F=0.10$\label{t:eigF10}} \tablehead { &
\multicolumn{2}{c}{$\BBeta=\infty$}&
             & \multicolumn{2}{c}{$\BBeta=100$}&
             & \multicolumn{2}{c}{$\BBeta=10$} \\
\cline{2-3}\cline{5-6}\cline{8-9}\\
  mode  & $\omgR/\Omega$& $\omgI/\Omega$ &&
          $\omgR/\Omega$& $\omgI/\Omega$ &&
          $\omgR/\Omega$& $\omgI/\Omega$}
\startdata 1 & 0.354 &$-6.363\times10^{-1}$&& 0.363
&$-7.086\times10^{-1}$&&
    0.379 &$-6.983\times10^{-1}$\\
2 & 0.809 &$-3.897\times10^{-1}$&& 0.817 &$-3.963\times10^{-1}$&&
    0.855 &$-4.454\times10^{-1}$\\
3 & 1.614 &$-1.078\times10^{-2}$&& 1.617 &$-1.173\times10^{-3}$&&
    1.619 &$-1.517\times10^{-2}$\\
4 & 2.740 &$+7.693\times10^{-3}$&& 2.741 &$+8.745\times10^{-3}$&&
    2.759 &$+8.427\times10^{-3}$\\
5 & 4.316 &$-3.565\times10^{-2}$&& 4.319 &$-2.545\times10^{-2}$&&
    4.377 &$-2.297\times10^{-2}$\\
6 & 6.019 &$-5.747\times10^{-2}$&& 6.052 &$-3.498\times10^{-2}$&&
    6.155 &$-3.220\times10^{-2}$\\
7 & 7.846 &$-5.831\times10^{-2}$&& 7.853 &$-3.778\times10^{-2}$&&
    7.996 &$-2.700\times10^{-2}$\\
8 & 9.663 &$-8.771\times10^{-2}$&& 9.678 &$-3.491\times10^{-2}$&&
    9.866 &$-2.868\times10^{-2}$\\
9 &11.509 &$-7.421\times10^{-2}$&&11.522 &$-3.069\times10^{-2}$&&
   11.752 &$-2.190\times10^{-2}$\\
10&13.352 &$-8.511\times10^{-2}$&&13.375 &$-2.640\times10^{-2}$&&
   13.645 &$-1.699\times10^{-2}$\\
\hline\\
 & \multicolumn{2}{c}{$\BBeta=5$} & & \multicolumn{2}{c}{$\BBeta=3$} &
 & \multicolumn{2}{c}{$\BBeta=1$ } \\
\cline{2-3}\cline{5-6}\cline{8-9}\\
  mode  & $\omgR/\Omega$& $\omgI/\Omega$ &&
          $\omgR/\Omega$& $\omgI/\Omega$ &&
          $\omgR/\Omega$& $\omgI/\Omega$\\
\hline\\
1 & 0.894 &$-5.006\times10^{-1}$&& 0.935 &$-5.721\times10^{-1}$&&
    1.033 &$-8.362\times10^{-1}$\\
2 & 1.622 &$-2.052\times10^{-2}$&& 1.624 &$-2.678\times10^{-2}$&&
    1.629 &$-5.097\times10^{-2}$\\
3 & 2.776 &$+8.532\times10^{-3}$&& 2.796 &$+8.896\times10^{-3}$&&
    2.877 &$+9.385\times10^{-3}$\\
4 & 4.433 &$-2.016\times10^{-2}$&& 4.498 &$-1.702\times10^{-2}$&&
    4.745 &$-7.109\times10^{-3}$\\
5 & 6.250 &$-2.726\times10^{-2}$&& 6.361 &$-2.199\times10^{-2}$&&
    6.776 &$-7.802\times10^{-3}$\\
6 & 8.132 &$-2.718\times10^{-2}$&& 8.286 &$-2.087\times10^{-2}$&&
    8.863 &$-2.309\times10^{-3}$\\
7 &10.042 &$-1.719\times10^{-2}$&&10.241 &$-1.584\times10^{-2}$&& - & - \\
8 &11.964 &$-1.949\times10^{-2}$&&12.207 &$-7.872\times10^{-3}$&& - & - \\
9 &13.896 &$-7.560\times10^{-3}$&&14.180 &$-2.576\times10^{-3}$&& - & - \\
10& - & - && - & - && - & - \\
\enddata
\end{deluxetable}

\clearpage

\begin{figure*}[!th]
\epsscale{0.8}\plotone{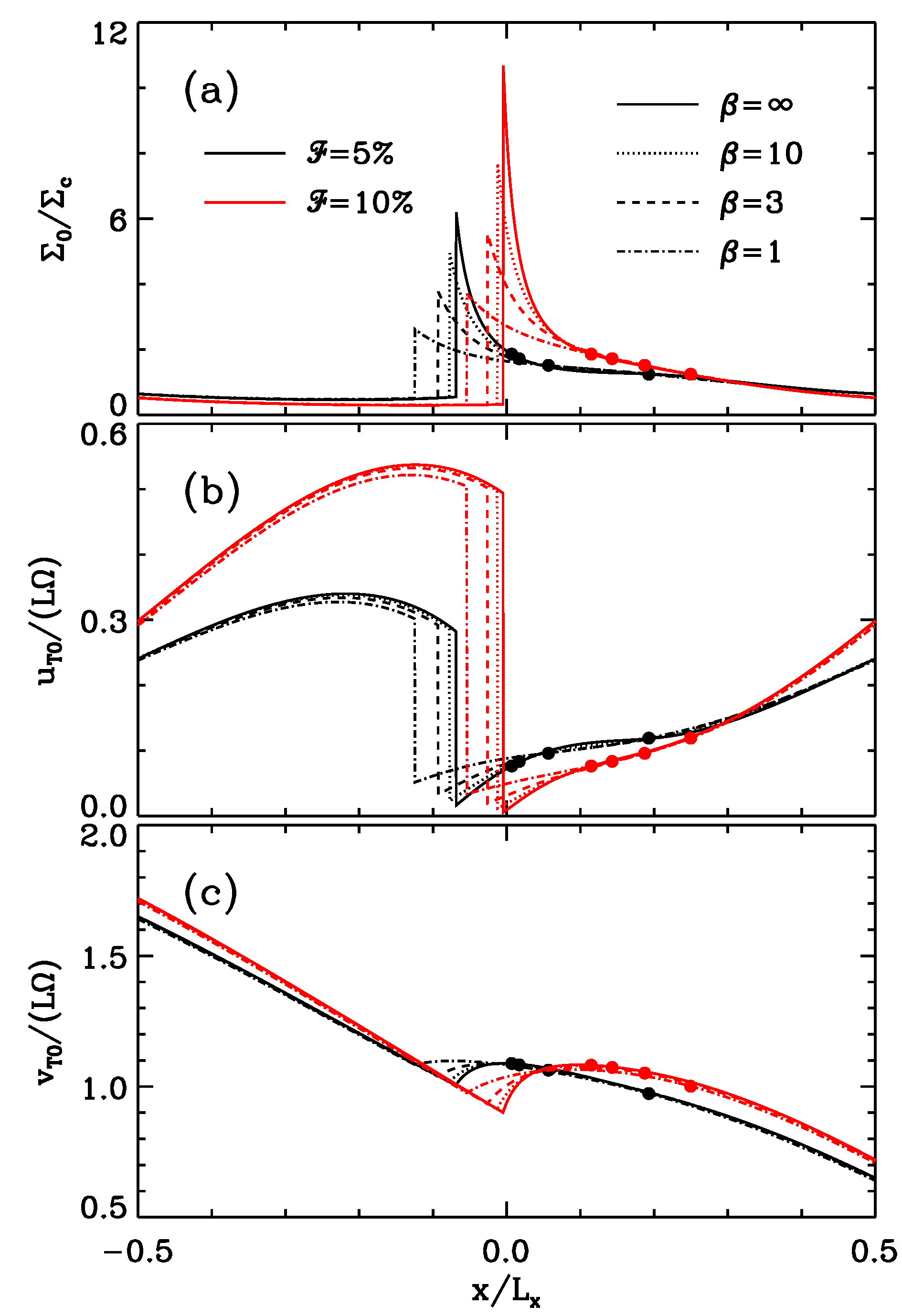} \caption{One-dimensional steady-state
shock profiles for $\F=5\%$ (black) and 10\% (red) with differing
$\BBeta$.  Each dot marks the magnetosonic point. The shock becomes
weaker for smaller $\F$ and $\BBeta$.  Note that shear reversal in the
immediate postshock regions is stronger for larger $\F$ and $\BBeta$.
\label{f:shock}}
\end{figure*}

\begin{figure*}[!th]
\epsscale{0.9}\plotone{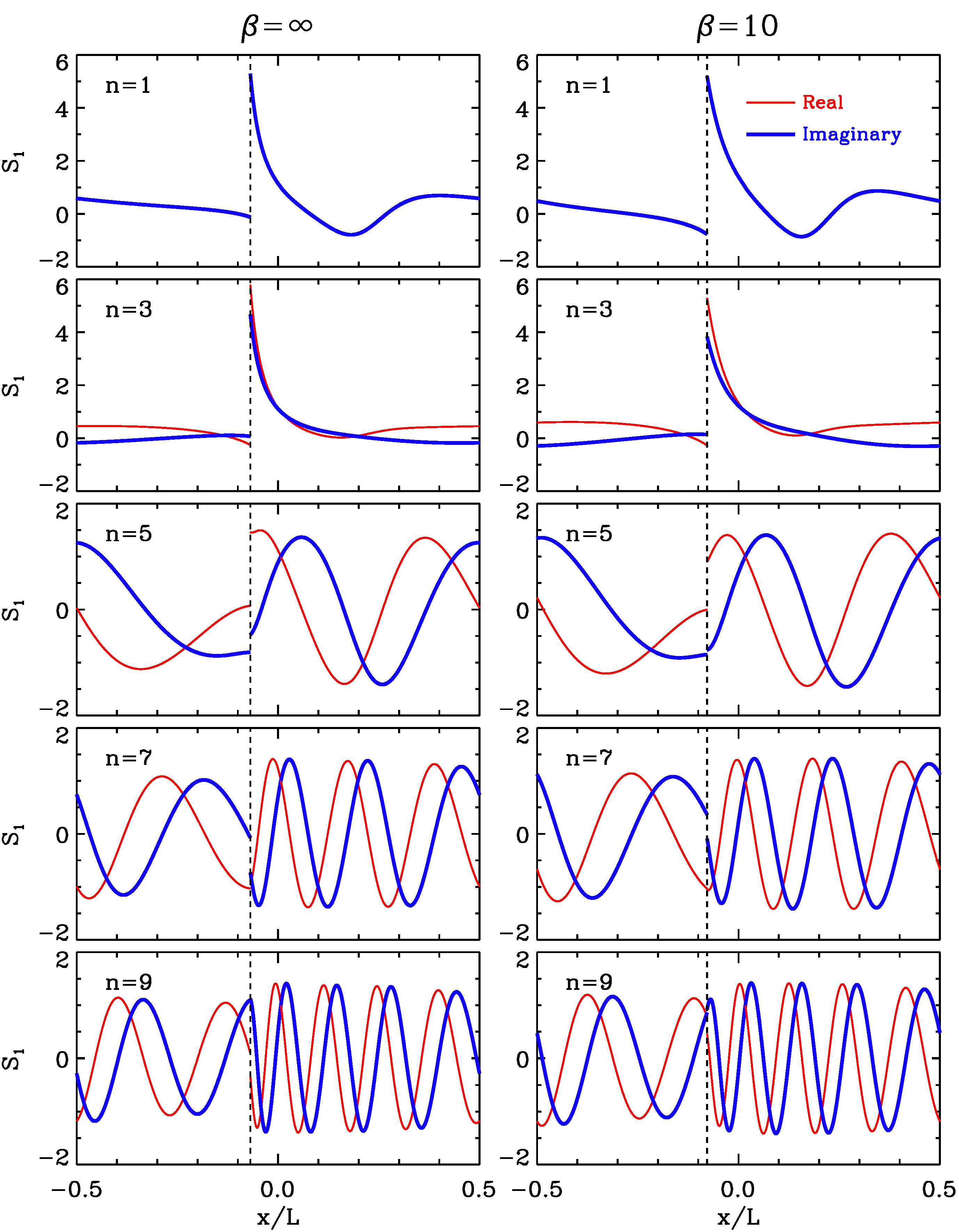} \caption{Five odd-mode eigenfunctions
$S_1$ for 1D perturbations with $\ky=0$ when (left) $\BBeta=\infty$ and
(right) $\BBeta=10$. The spiral forcing is set to $\F=5\%$. Red and
blue lines represent the real and imaginary parts, respectively. All
values are normalized such that $\text{Re}(S_1)=\text{Im}(S_1)=1$ at
the magnetosonic point located at $x/L=0.007$ and $x/L=0.017$ for
$\BBeta=\infty$ and 10, respectively. The vertical line in each panel
indicates the shock front. \label{f:1eigfun}}
\end{figure*}

\begin{figure*}[!th]
\epsscale{0.9}\plotone{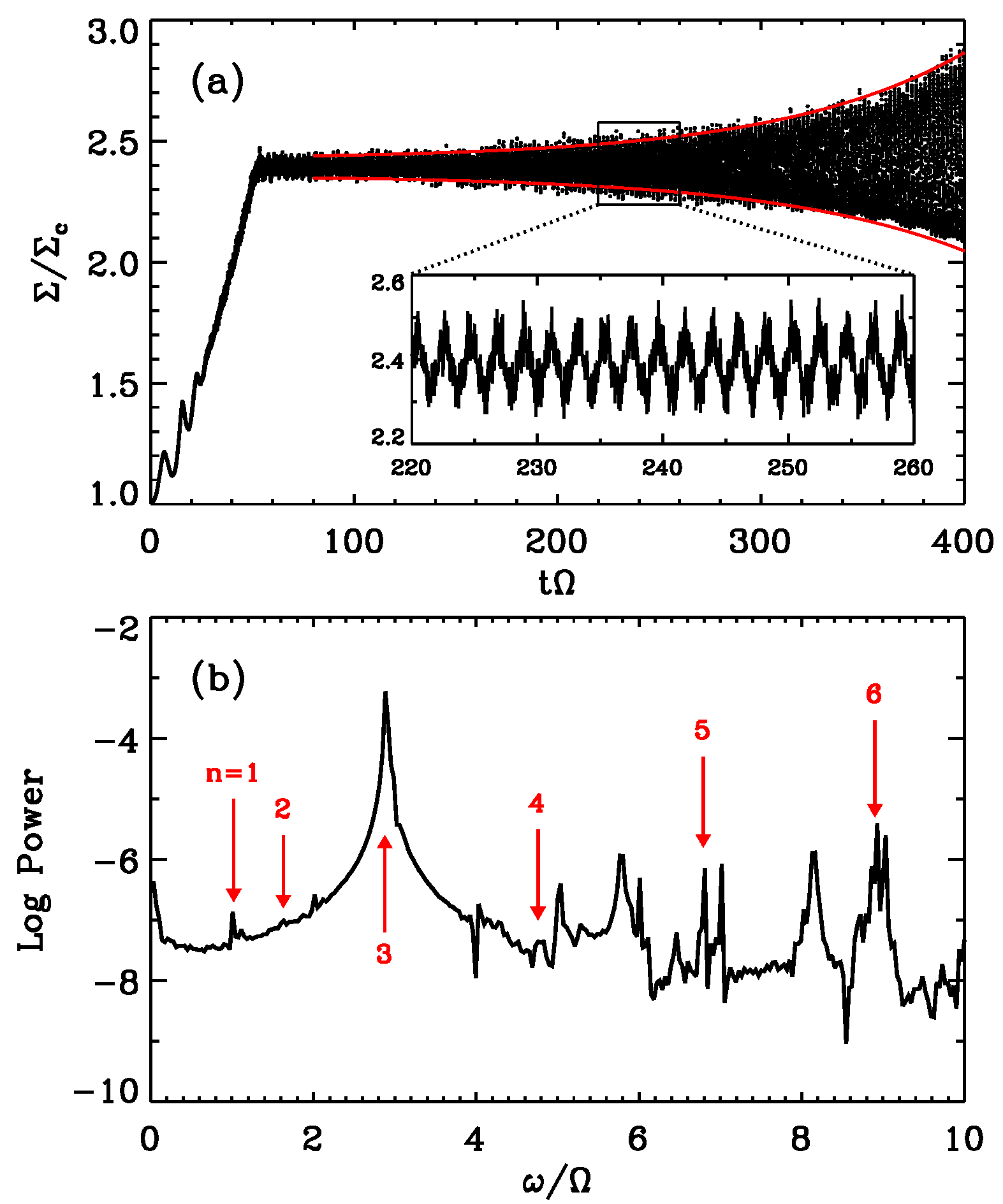} \caption{(a) Temporal variations of
the gas surface density $\Sigma$ at $x=0$ from a 1D simulation with
$\F=10\%$ and $\BBeta=1$.  Red solid lines envelope the fluctuation
amplitudes of $\Sigma$.  The inset zooms in the time range $220\leq
t\Omega \leq 260$ to display the density fluctuations. (b) The power
spectrum of the density fluctuations. The frequencies marked by the red
arrows represent the real parts of the eigenvalues listed in Table
\ref{t:eigF10}.  \label{f:1dsim}}
\end{figure*}

\begin{figure*}[!th]
\epsscale{0.9}\plotone{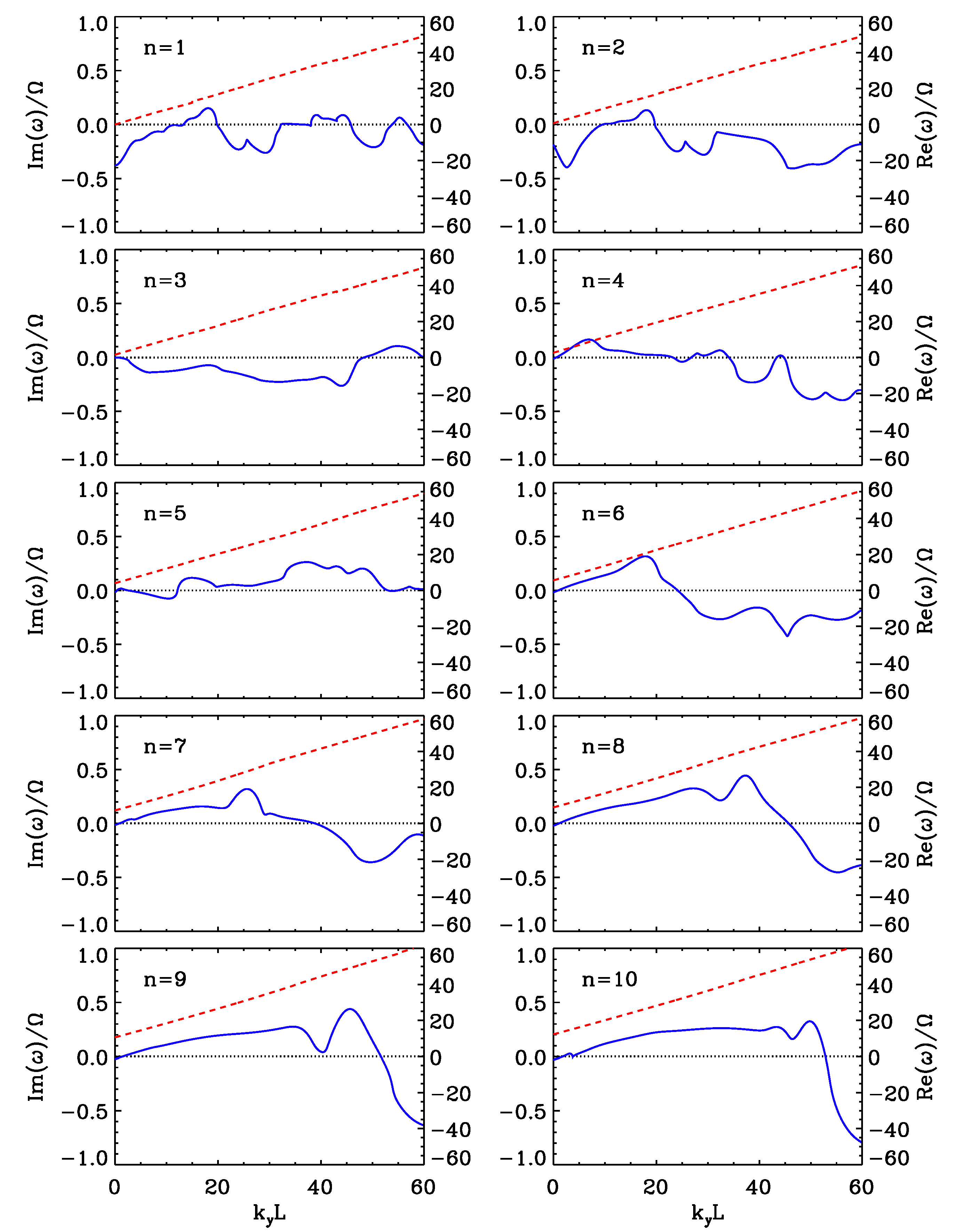} \caption{Non-axisymmetric dispersion
relations of the ten lowest-frequency eigenmodes for $\F=5\%$ and
$\BBeta=10$. The modes are numbered in the increasing order of $\omgR$
at $\ky=0$. In each panel, the blue solid line (left $y$-axis) gives
$\omgI$, while the red dashed line (right $y$-axis) is for $\omgR$.
\label{f:disp2d}}
\end{figure*}

\begin{figure*}[!th]
\epsscale{0.9}\plotone{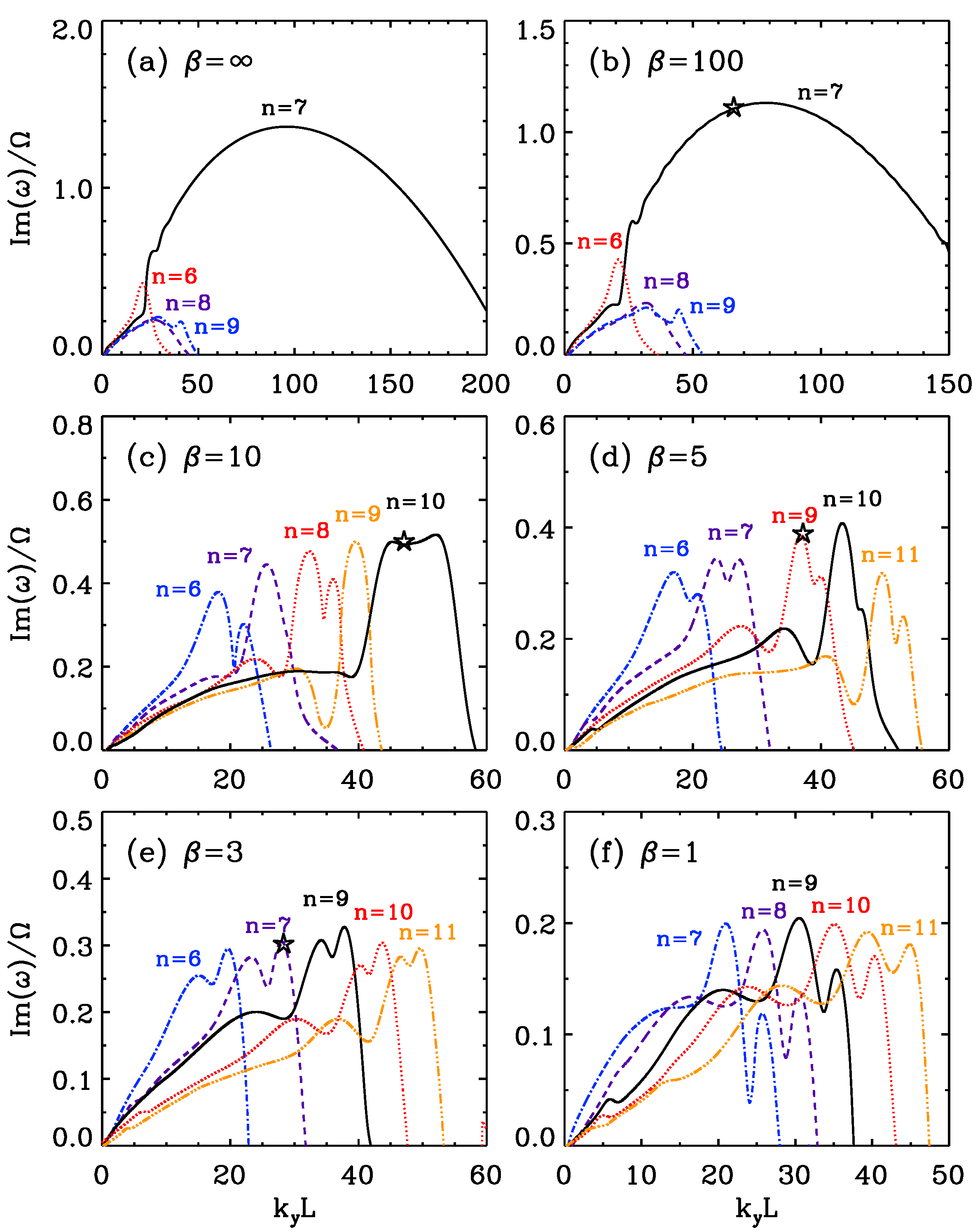} \caption{Dependence on $\BBeta$ and
$\ky$ of the growth rates $\omgI$ of the dominant overstable modes for
$\F=5\%$. For $\BBeta\geq100$, the WI is dominated by the $n=7$ mode,
while several modes shown have similar growth rates for $\BBeta\leq10$.
Star symbols mark the growth rates and wavelengths of the WI measured
from direct numerical simulations in Section \ref{sec:num}.
\label{f:disp2dF05}}
\end{figure*}

\begin{figure*}[!th]
\epsscale{0.9}\plotone{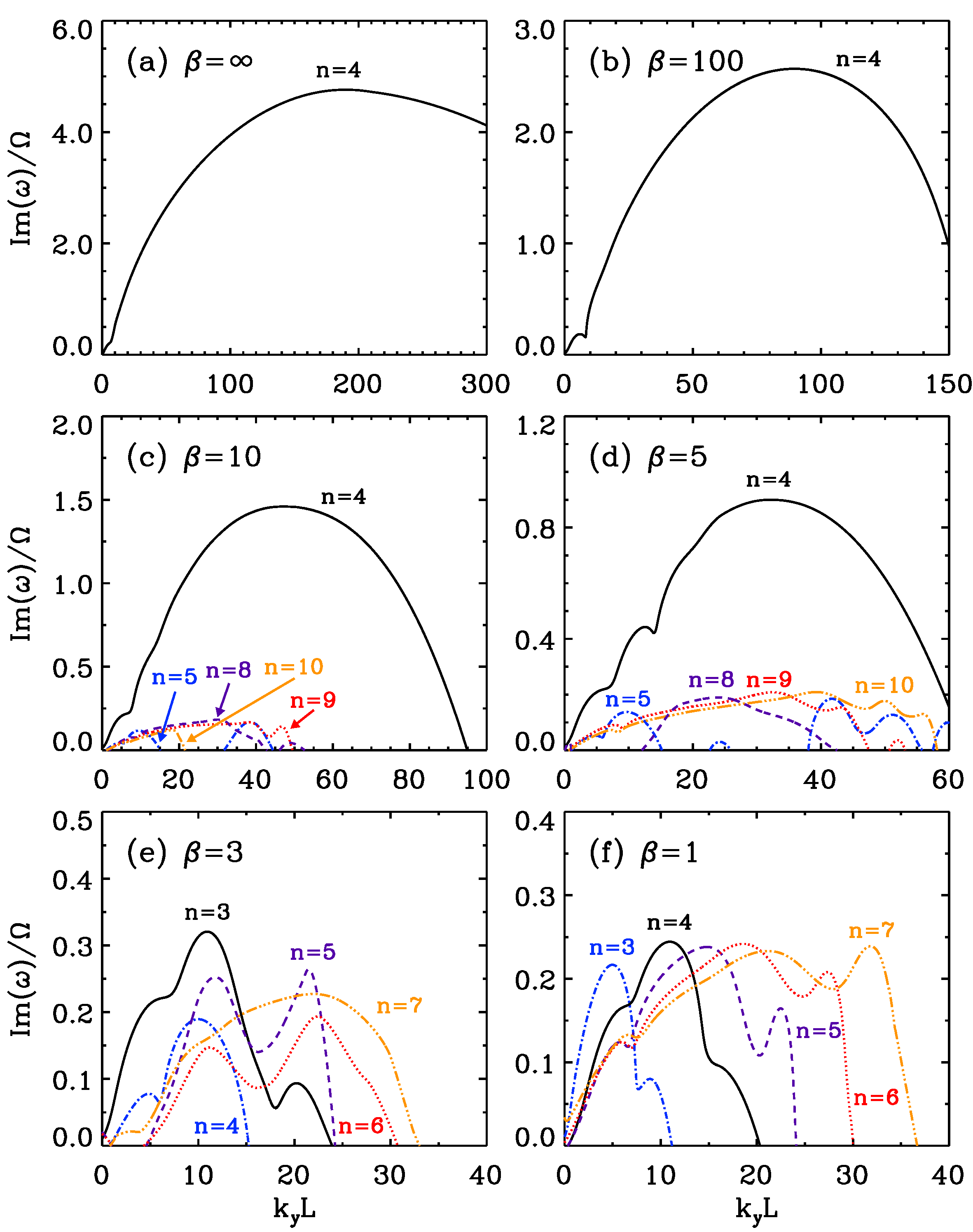} \caption{Dependence on $\BBeta$ and
$\ky$ of the growth rates $\omgI$ of the dominant overstable modes for
$\F=10\%$. For $\BBeta\geq5$, the WI is dominated by the $n=4$ mode,
while several different modes have similar growth rates for
$\BBeta\leq3$. \label{f:disp2dF10}}
\end{figure*}

\begin{figure*}[!th]
\epsscale{1.0}\plotone{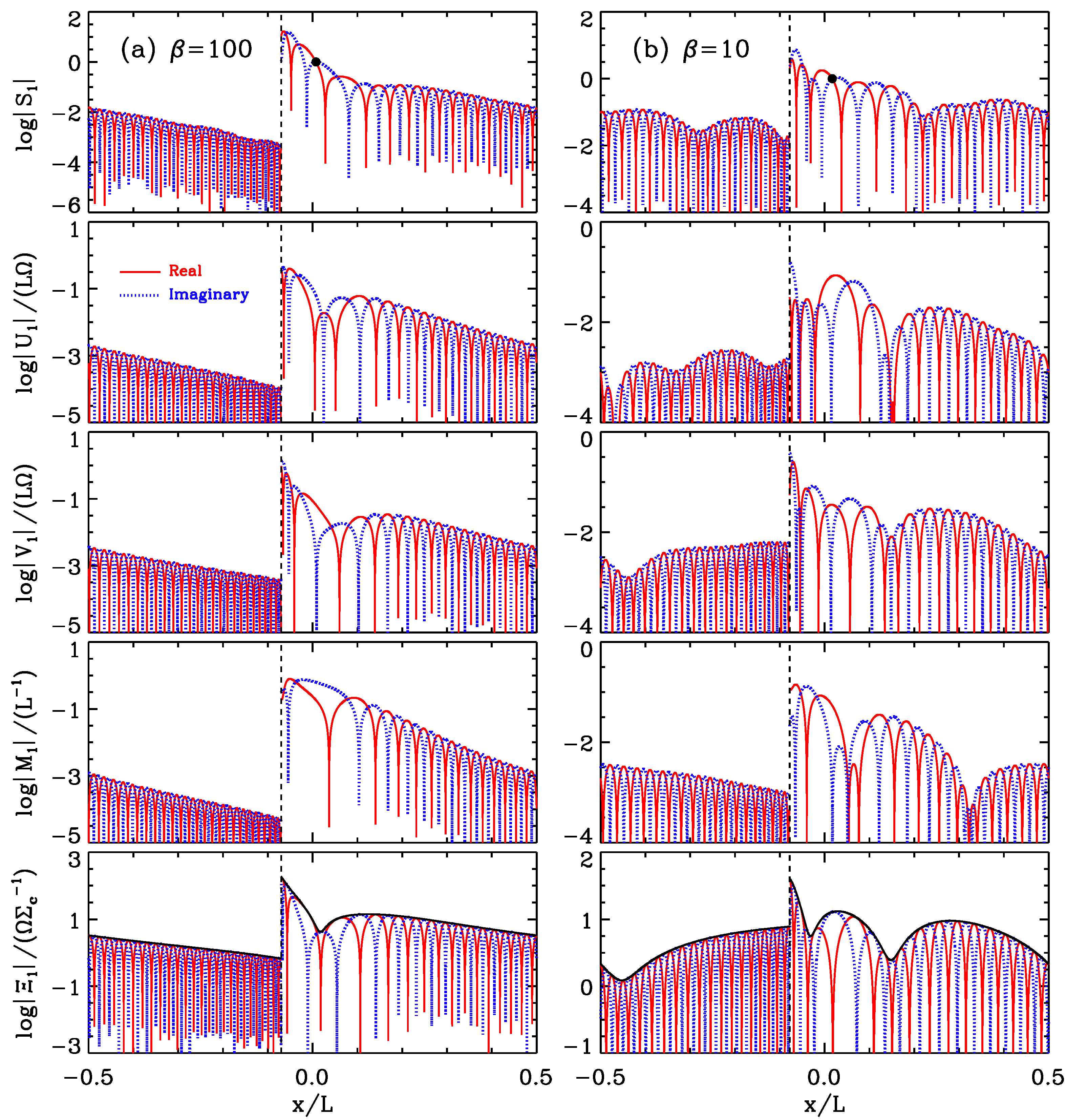}\caption{Eigenfunctions $S_1$, $U_1$,
$V_1$, $M_1$, and $\Xi_1$ of an overstable mode (left) with
$\omega/\Omega=84.4 + 1.13i$ and $\ky L=78.1$ for $\BBeta=100$ and
$\F=5\%$ and (right) with $\omega/\Omega=59.1+0.50i$ and $\ky L=53.4$
for $\BBeta=10$ and $\F=5\%$. The red solid and blue dotted curves
represent the absolute values of the real and imaginary parts. The
shock front is indicated by the vertical dashed line in each panel. The
black dots in the top panels mark the magnetosonic points. The
black solid lines enveloping the eigenfunctions in the bottom panels
draw the solutions of Equation \eqref{e:intpv}. \label{f:2deigen}}
\end{figure*}

\begin{figure*}[!th]
\epsscale{1.0}\plotone{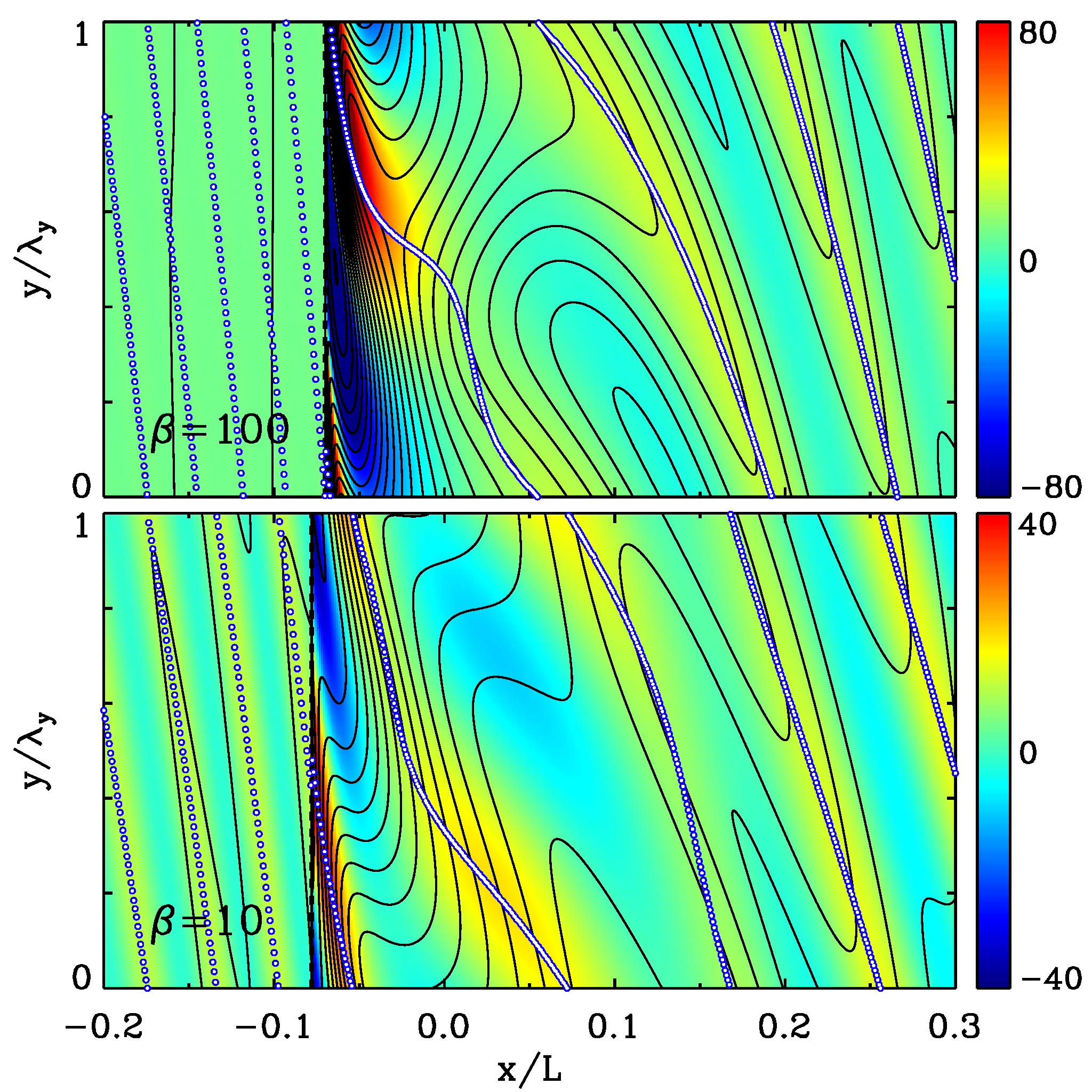} \caption{Distributions of the
perturbed magnetic fields (black solid lines) in the regions with
$-0.2\leq x/L \leq 0.3$ and $0 \leq y/\lambda_y \leq 1$ overlaid over
the perturbed PV, $\text{Re}(\xi_1)$, displayed in color scale for the
(a) $\BBeta=100$ and (b) $\BBeta=10$ cases with $\F=5\%$. White dots in
both panels trace the wavefronts of the perturbed PV. Colorbar labels
$\text{Re}(\xi_1)/(\Omega\Sigma_c^{-1})$. \label{f:2deigenmap}}
\end{figure*}

\begin{figure*}[!th]
\epsscale{1.0}\plotone{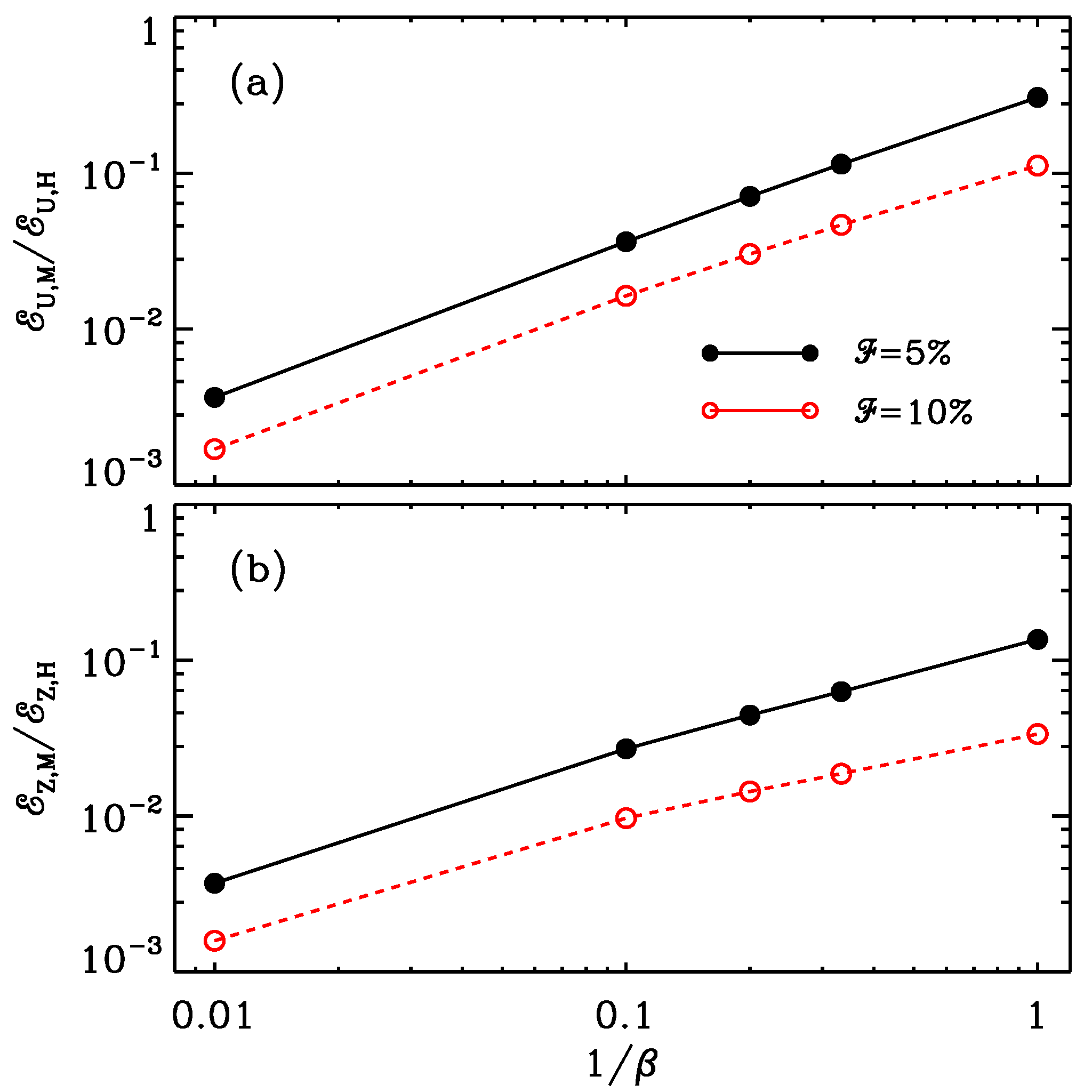} \caption{Ratio of the magnetic to
hydrodynamic terms in the expression for the PV jumps across a
disturbed shock front due to (a) the perpendicular velocity $U_1$ and
(b) the distortion amplitude $Z_1$. Filled and open circles correspond
to $\F=5\%$ and 10\%, respectively. \label{f:coeff}}
\end{figure*}

\begin{figure*}[!th]
\epsscale{1.0}\plotone{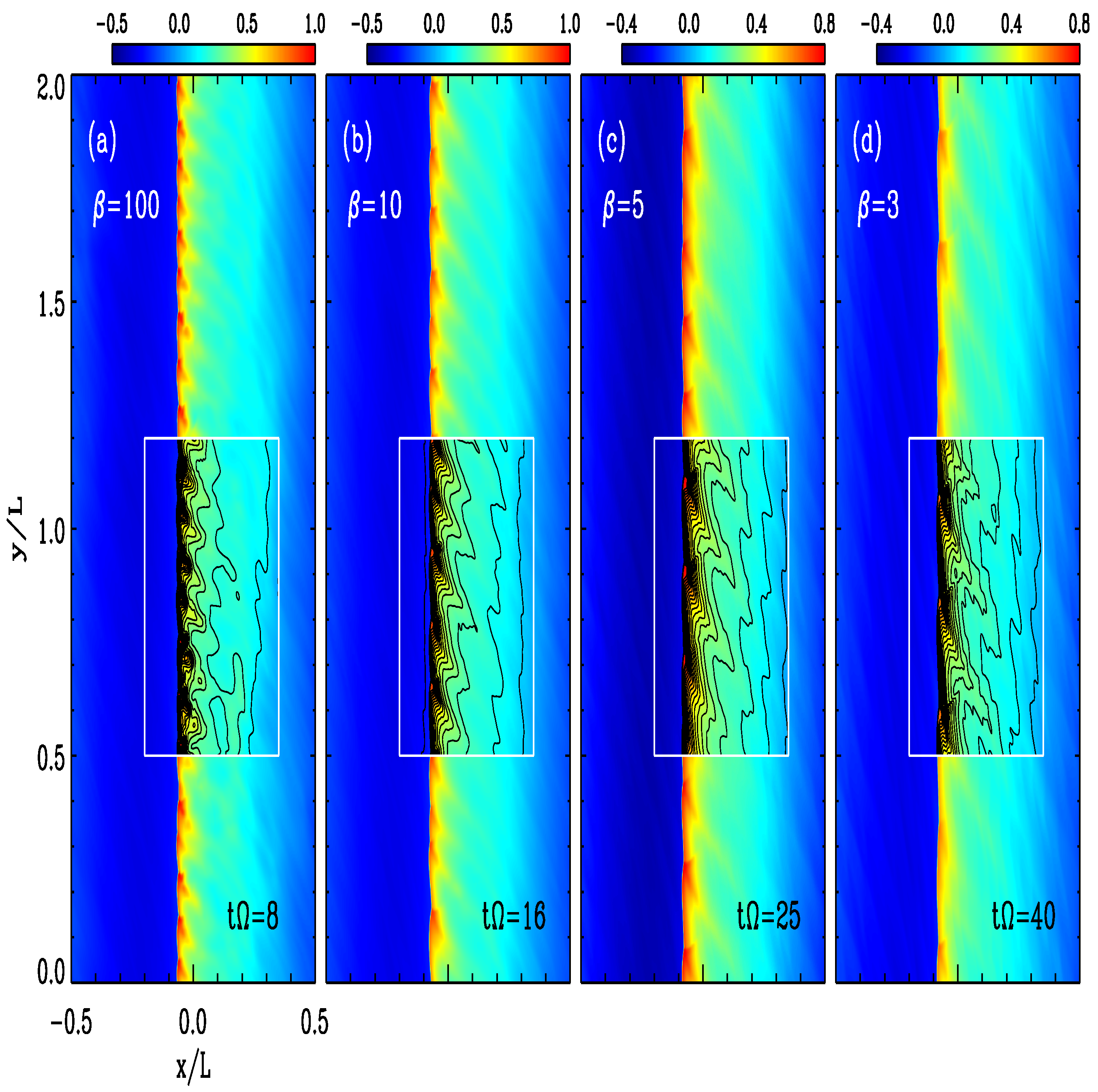} \caption{Snapshots of the gas surface
density as well as the configuration of the magnetic fields in the
regions with $-0.2\leq x/L \leq 0.35$ and $0.5\leq y/L\leq 1.2$ from 2D
simulations with $\BBeta=100$, 10, 5, and 3 at $t\Omega=8$, 16, 25, and
40 from left to right, respectively. The arm forcing is $\F=5\%$ and
the grid resolution over the domain size of $L\times 2L$ is $2048\times
4098$. The number of the nonlinear features grown by the WI along the
$y$-direction is 21, 15, 12, and 9 from left to right. Colorbar labels
$\log(\Sigma/\Sigma_c)$. \label{f:2dsim}}
\end{figure*}

\begin{figure*}[!th]
\epsscale{1.0}\plotone{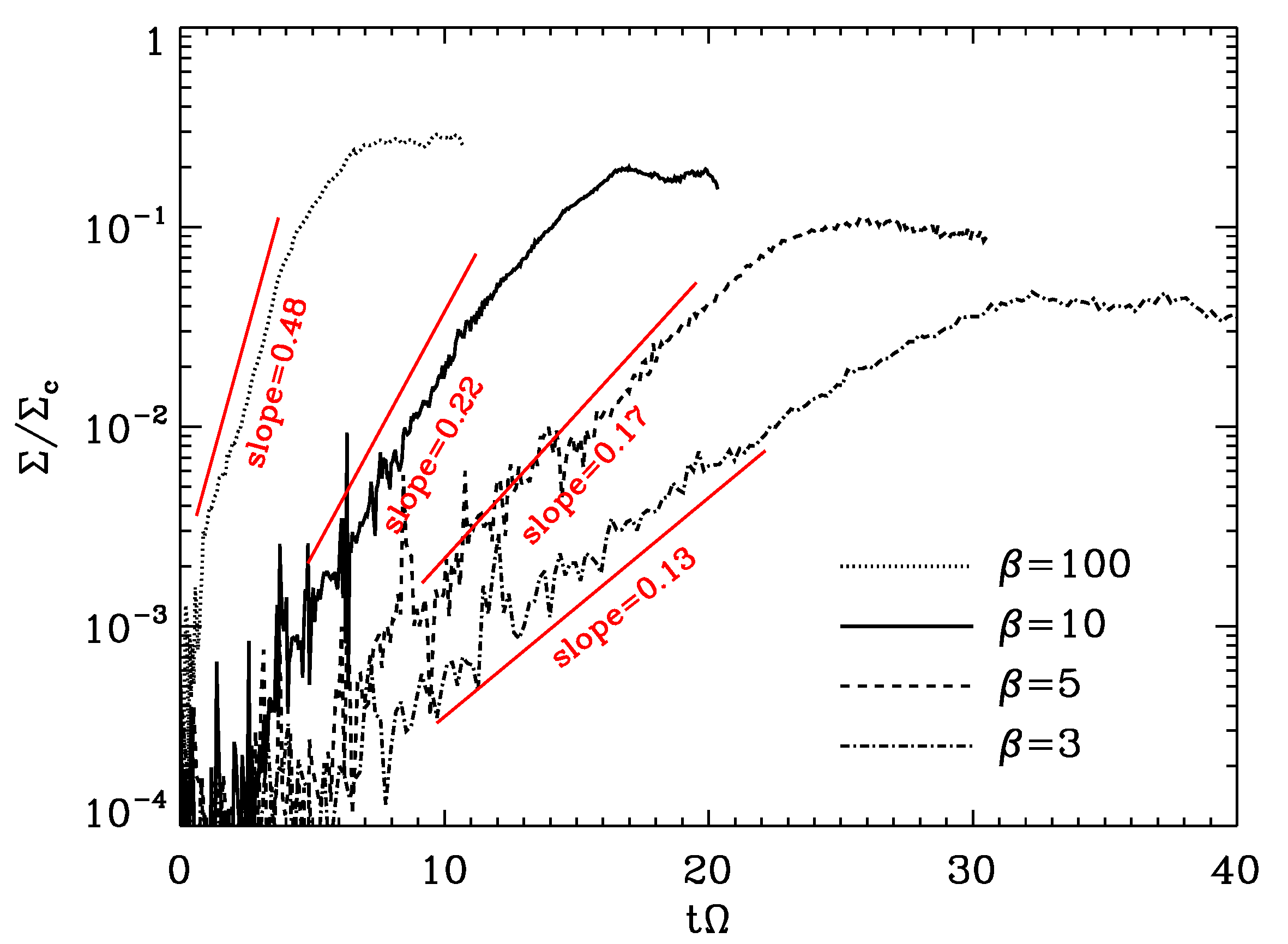} \caption{Temporal evolution of the
maximum surface density measured at $x=0$ from the models shown in
Figure \ref{f:2dsim}. The growth rate measured from the slope indicated
as the line segment in each model is  consistent with the results of
the normal-mode linear stability analysis. \label{f:growth}}
\end{figure*}

\begin{figure*}[!th]
\epsscale{1.0}\plotone{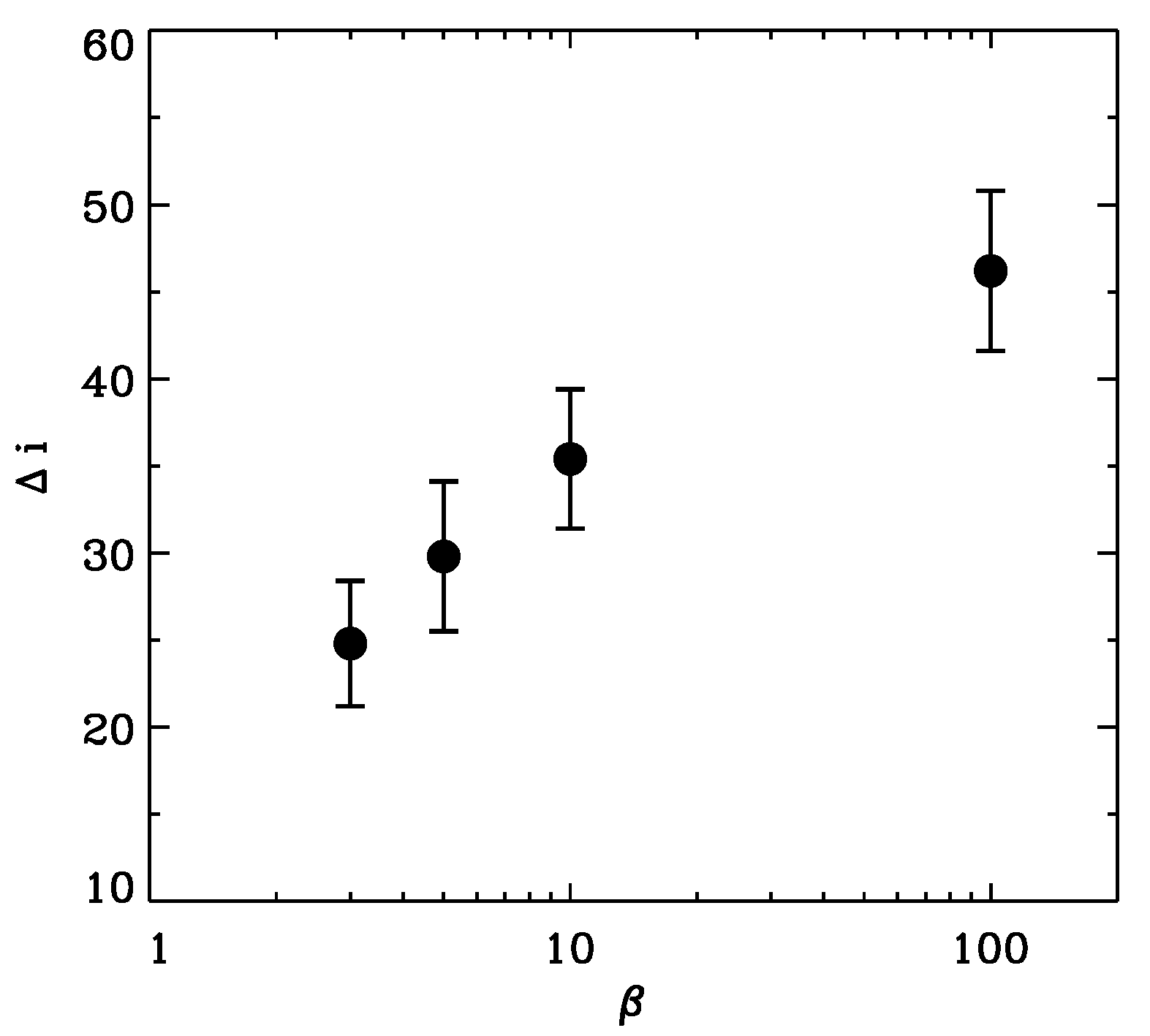} \caption{Difference between the pitch
angles of the main arm and the nonlinear structures stretched from it
in the numerical simulations with $\F=5\%$, averaged over the regions
with $\xSH \leq x \leq \xSH+L/2$.  Filled circles and errorbars give
the mean values and standard deviations over the time interval of
$\Delta t=5/\Omega$ from the time epoch shown in Figure \ref{f:2dsim}.
\label{f:pitch}}
\end{figure*}

\end{document}